# When Deep Learning Meets Digital Image Correlation

S. Boukhtache[1], K. Abdelouahab[2], F. Berry[1], B. Blaysat[1], M. Grédiac[1], F. Sur[3]

[1]*Institut Pascal, UMR 6602, Université Clermont-Auvergne, CNRS, SIGMA Clermont, Clermont-Ferrand, France*

[2]*Sma-RTy SAS, Aubière, France*

[3]*LORIA, UMR 7503, Université de Lorraine, CNRS, Inria, Nancy, France*

## 1. Introduction

Digital Image Correlation (DIC) is a full-field displacement and strain measurement technique which has rapidly spread in the experimental mechanics community. The main reason is that this technique achieves a good compromise between versatility, ease of use, and metrological performance [1]. Many recent papers illustrate the use of DIC in various situations [2]. Some others discuss how to characterize or improve its metrological performance, as [3] written within the framework of the DIC Challenge [4]. Despite all its advantages, DIC suffers from some drawbacks. For instance, DIC is by essence an iterative procedure, which automatically leads to mobilizing significant computational resources. Consequently, its use is limited when dense (i.e. pixelwise-defined) displacement or strain distributions are to be measured. Another drawback is the fact that DIC acts as a low-pass filter, which causes the retrieved displacement and strain fields to be blurred. Indeed, it is shown in [5, 6] that the displacement field rendered by a DIC system is not the actual one, but regardless of noise, the actual one convolved by a Savitzky-Golay filter. This makes DIC to be unable to correctly render displacement or strain fields featuring high spatial frequencies. Overcoming these limitations seems however difficult without introducing a paradigm shift in image processing itself. For instance, the minimization of the optical residual that DIC performs iteratively in the spatial domain can be switched to the Fourier domain [7]. The benefit is to considerably reduce the computing time and to allow the use of optimal patterns in terms of camera sensor noise propagation [8, 9, 10]. The drawback is that such patterns are periodic, and depositing them on specimens remains challenging as long as no simple transferring technique is commercially available. In the present study, we propose to use random speckle patterns as in DIC and to investigate to what extent a Convolutional Neural Network (CNN) can be used to retrieve displacement and strain fields from a pair of reference and deformed images of a speckle pattern. To the best of the authors' knowledge, this route has never been explored so far to retrieve dense subpixel displacement fields. However, CNNs have been widely used in computer vision in the recent past, but mainly in order to perform image classification or recognition [11], or to estimate rigid-body displacements of solids [12]. The problem addressed here is somewhat different because deformation occurs, and because the resulting displacement is generally much lower in amplitude than in the aforementioned cases of rigid-body movements. Consequently we mainly focus here on the estimation of subpixel displacements.

The present paper is organized as follows. The basics of CNNs and deep learning are first briefly given in Section 2. We examine in Section 3 how a problem similar to ours, namely optical flow determination, has been tackled with CNNs in the literature. CNNs must be trained, and since no dataset suited to the problem at hand is available in the literature, we explain in Section 4 how a first dataset containing speckle images has been generated. Then we test in Section 5 how four pre-existing networks are fine-tuned with this dataset, and examine whether these networks are able to determine displacement fields from different speckle images artificially deformed. The network giving the best results is then selected and improved in several ways to give displacement fields of better quality. The different improvements and the corresponding results are given in Section 6. Finally, we use the network resulting from all the suggested improvements, named here "StrainNet", to process some pairs of speckle images from the DIC Challenge and from a previous study.

The numerical experiments proposed in this paper can be reproduced with Matlab and Pytorch codes as well as with datasets available at the following URL: https://github.com/DreamIP/StrainNet.

## 2. A short primer on deep learning

Data-driven approaches have revolutionized several scientific domains in the last decade. This is probably due to a combination of several factors, in particular the dramatic improvement of computing and information storage capacity. For example, it is now quite easy to have large datasets gathering many examples for a task of interest, as millions of labeled images for an image classification task. However, the most crucial point is probably the rise of new machine learning techniques built on artificial neural networks. While neural networks have been introduced in the 1950s and have been continuously improved since then, a major breakthrough occurred a few years ago with the demonstration that deep neural networks [13] give excellent results in many signal processing applications. The most iconic breakthrough is probably the famous 2012 paper [11] which demonstrates a deep learning based system outperforming the competing approaches in the ImageNet classification challenge.

The basic elements of neural networks are neurons connected together. In feedforward neural networks, neurons are organized in layers. The input layer encodes input data (an image in image classification tasks, two images in the present problem) and the output layer encodes the associated label (the posterior probability of a class in classification, a displacement field here). The intermediate layers are the hidden layers. They are made of neurons which output a signal. Each neuron of the hidden and output layers are connected to neurons from the preceding layer. The output of these neurons are a weighted sum of the connected neurons modulated by a continuous non-decreasing function called activation function, except for the output layer in a regression problem as here where no activation is involved. The most popular activation function is the so-called Rectified Linear Unit (ReLU) [13].

Deep neural networks are called "deep" because they may be made of several tens of layers. As we have seen, neural connections between layers involve weights. Note that "weights" are also commonly referred to as the "network parameters" in the literature. The term "weight" is used here to avoid confusion with the other parameters defined in the paper.

Computer vision applications typically call for a subclass of deep neural networks known as convolutional neural networks (CNNs) where the number of weights is mitigated by imposing that the weighted sums correspond to convolutions, which turn out to be the basic operators of signal processing. Neurons from a convolutional layer are represented as the output of the implemented convolutions, hence the blue parallelepipeds in Figure 1. Another ingredient used in CNN is down-sampling which reduces the width of the layers involved. In the present work, down-sampling by a factor (the so-called stride) of two is obtained by computing convolutions shifted by two units instead of one unit as in a standard convolution. This explains the narrower and narrower layers in the feature extraction part in Figure 1. Note that in this figure, up-sampling is performed through the so-called transposed convolutions when predicting the displacement field.

However, deep CNNs are still likely to require several millions of weights. As in any supervised learning task, the weights are set by training the CNN over a dataset made of observations, that is, pairs of inputs and corresponding expected outputs. More precisely, training, called deep learning in this context, consists in minimizing with respect to the weights the cost of errors between the expected outputs and the CNN outputs obtained from the inputs and the current weights. The error cost is measured through the so-called loss function. The optimization method implementing the training process is most of the time a variation of the mini-batch stochastic gradient descent (SGD) algorithm: small sets of observations are iteratively randomly drawn from the whole dataset, giving a so-called mini-batch, and the weights of the CNN are then slightly updated in order that the sum of the losses over each mini-batch decreases. At each iteration, the outputs of the CNN computed from the mini-batch inputs are thus supposed to get closer to the expected outputs. The magnitude of the weight update is proportional to the step size of the SGD algorithm, called learning rate in machine learning applications and denoted $\lambda$ in this paper. This parameter has to be set carefully. Each complete pass over the dataset is called an epoch. Many epochs are needed to perform the full training of a CNN.

However, training a deep CNN requires a large dataset and heavy computational resources, which is simply not possible in many situations. A popular alternative approach consists in using freely available pre-trained networks, that is, CNN that have been trained on standard datasets, and in adjusting their weights to the problem of interest. Adjustment can be performed by fine tuning or transfer learning. The former consists in marginally changing the weights in order to adapt them to the problem of interest and its dataset. The latter consists in changing a part of the architecture of the pre-trained network and in learning the corresponding weights from the problem of interest, the remaining weights being kept constant. Let us now examine how such networks have been used in the literature for solving a classic computer vision problem similar to ours, namely optical flow estimation.

## 3. A brief review of CNN-based methods for optical flow estimation

Optical flow is the apparent displacement field obtained from two views of a scene. It is caused by the relative motion of the observer and objects in the scene which may move or deform. Recent algorithms of optical flow estimation based on CNNs provide a promising alternative to the methods classically used to resolve this problem in computer vision after the seminal papers by Horn and Schunck [14] or by Lucas and Kanade [15]. As a typical machine learning setup, AI-based optical flow estimation algorithms can be divided into three categories: unsupervised, semi-supervised, or supervised. Unsupervised [16, 17, 18] and semi-supervised [19, 20] methods are reviewed in the literature to address the problem of limited training data in optical flow estimation. In contrast, these methods do not yet reach the accuracy of their supervised counterparts. Furthermore, supervised methods are the predominant way of learning, as described in the preceding section, and generally provide good performance. However, they require a large amount of accurate, ground-truth optical flow measurements for training. Most accurate models use CNNs as one of the components of their system, as in DCFlow [21], MRFlow [22], Deepflow [23], and Flow Fields [24]. None of these previous approaches provide end-to-end trainable models or real-time processing performance. The most efficient algorithms recently proposed in the literature for optical flow estimation are reviewed below. In general, they share the same architecture (a schematic view is represented in Figure 1). The first part of the network extracts the features of the images and the optical flow is predicted through an up-sampling process in the second part of the network.

Dosovitskiy *et al.* [12] presented two CNNs called FlowNetS and FlowNetC to learn the optical flow from synthetic dataset. These two CNNs are constructed based on the U-Net architecture [25]. FlowNetS is an end-to-end CNN-based network. It concatenates the reference and the current images together and feeds them to the network in order to extract the optical flow. In contrast, FlowNetC creates two separate processing pipelines for each of these two images. Then, it combines them with a correlation layer that performs multiplicative patch comparisons between the two generated feature maps. The resolution of the output optical flow in both networks is equal to 1/4 of the image resolution. Indeed, the authors explain that adding other layers to reach full resolution is computationally expensive, and does not really improve the accuracy. A consequence is that with this network, the optical flow at full resolution is obtained by performing a bilinear interpolation with an upscale factor of 4. This point is discussed further below as we do not use exactly the same approach to solve the problem addressed in this paper.

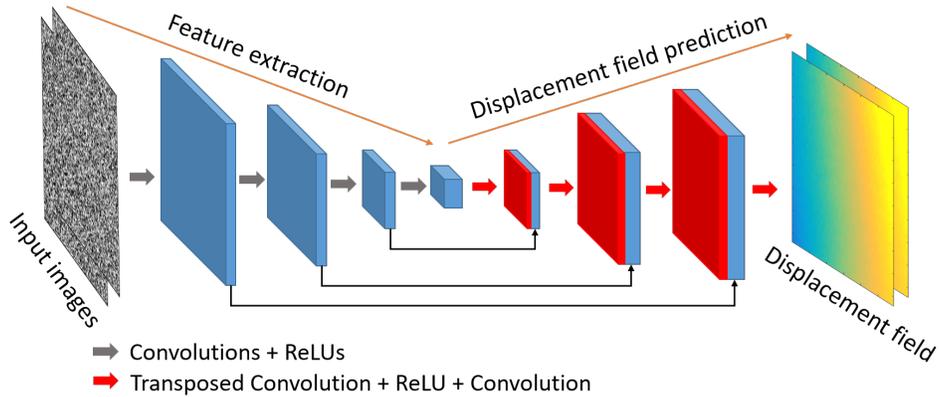

Figure 1: Schematic view of a convolutional neural network. Anticipating the results discussed in Section 6.1, we present here a schematic view of FlowNet-f. The feature extraction part consists of several convolution layers followed by a ReLU (Rectified Linear Unit [13]). The last layer has a stride of two to perform down-sampling. Such a stack is called a level. The output of these levels can thus be represented by narrower and narrower blocks, in blue here. In the displacement prediction part, the levels are made of transposed convolution layers (for up-sampling) and convolution layers. The input of each level of this part is the output of the preceding level, concatenated to the output of the levels from the feature extraction part, as represented by the black arrows.

Hu *et al.* [26] proposed a recurrent spatial pyramid network that inputs the full-resolution images and generates an initial optical flow of 1/8 full-resolution. The initial flow is then upscaled and refined recurrently based on an energy function. The initial generated optical flow is converted to reach the full-resolution flow by performing this recurrent operation on the spatial pyramid. This network achieves comparable performance to FlowNetS with 90 times less weights. It is however twice as slow as FlowNetS because of the recurrent operations performed on the spatial pyramid. Note that when CNNs are compared, it is important to use the same image size and the same GPU for all of them. In addition, the computing time, or inference time, is generally considered as equal to the mean time value over a certain number of pairs of images.

FlowNet 2.0 [27] is the extension of FlowNet [12]. It stacks multiple FlowNetS and FlowNetC networks in order to obtain a more accurate model. FlowNet 2.0 reduces the estimation error by more than 50% compared to FlowNet. However it has three limitations. First, the model size is 4 times larger than the original FlowNet (over 160 million weights). Second, FlowNet 2.0 is 4 times slower than FlowNet. Third, the sub-networks need to be trained sequentially to reduce overfitting problem. Overfitting is a non-desirable phenomenon, which occurs in deep learning when the cardinality of the dataset is too small with respect to the number of weights to be trained. This leads to a network, which fails in correctly predicting the optical flow from images which are too different from those contained in the training dataset.

Ranjan *et al.* [28] developed a compact spatial pyramid network, called SpyNet. It combines deep learning with classical optical flow estimation principle such as image warping at each pyramid level, in order to reduce the model size. In SpyNet, the pyramid levels are trained independently. Then, each level uses the previous one as an initialization to warp the second image toward the first through the spatial pyramid network. It achieves similar performance to FlowNetC on the same benchmark but is not as accurate as FlowNetS and FlowNet2. In contrast to FlowNet [12], SpyNet model is about 26 times smaller but almost similar in terms of running speed. In addition, it is trained by level, which means that it is more difficult to train than FlowNet.

LiteFlowNet [29] is composed of two compact sub-networks specialized in pyramidal feature extraction and optical flow estimation, respectively named NetC and NetE. NetC transforms the two input images into two pyramids of multi-scale high-dimensional features. NetE consists of cascaded "flow-inference" and "regularization" modules that estimate the optical flow fields. LiteFlowNet performs better than SpyNet and FlowNet, but it is outperformed by FlowNet 2.0 in terms of accuracy. It has 30 times less weights than FlowNet 2.0, but it is 1.36 times faster.

Sun *et al.* [30] proposed PWC-Net. This network is based on simple and well-established principles such as pyramidal processing, warping, and cost volume. Similarly to FlowNet 2.0, PWC-Net shows the potential of combining deep learning and optical flow knowledge. Compared to FlowNet 2.0, the model size of PWC-Net is about 17 times smaller and the inference time is divided by two. This network is also easier to train than SpyNet [28]. PWC-Net outperforms all the previous optical flow methods on the MPI Sintel [31] and KITTI 2015 [32] benchmarks.

All these studies demonstrate the usability of CNNs for estimating the optical flow. This motivated the development of similar models in order to tackle analogous problems arising in other research fields. This is typically the case in [33], where the authors used FlowNetS to extract dense velocity fields from particle images of fluid flows. The method introduced in this reference provides promising results, but the experiments indicate that the deep learning model does not outperform classical methods in terms of accuracy. Cai *et al.* [34] developed a network based on the LiteFlowNet [29] to extract dense velocity fields from a pair of particle images. The original LiteFlowNet network is enhanced in order to improve the performance by adding more layers. This leads to more accurate results than those given in [33] at the price of extra computing time.

In the following section, we propose to employ a deep convolutional neural network to measure full-field displacement fields that occur on the surface of deformed solids. As for 2D DIC, we assume here that the deformation is plane and that the surface is patterned with a random speckle. The main difference with the examples from the literature discussed above is twofold. First a deformation occurs and second, the resulting displacements are generally small, which means that subpixel resolution must be achieved. In addition, the idea is to use this measurement in the context of metrology, meaning that the errors affecting these measurement fields must be thoroughly estimated. A specific dataset and a dedicated network christened "StrainNet" are proposed herein to reach these goals, as described in the following sections.

## 4. Dataset

### 4.1. Existing datasets

Supervised training of deep CNNs requires large datasets. In the context of optical flow, datasets are made of pairs of images together with a ground truth optical flow, which may be achieved by rendering synthetic images. In spite that generating such large datasets is a tedious task, several datasets were generated and used in previous studies on optical flow estimation [35]. The most commonly datasets used for optical flow estimation are listed in Table 1.

| Dataset | Number of frames | Resolution | Displacements |
|---|---|---|---|
| Middlebury [36] | 72 | varies | Small ($\leq$ 10 pixels) |
| KITTI2012 [37] | 194 | 1226×370 | Large |
| KITTI2015 [32] | 800 | 1242×375 | Large |
| MPI Sintel [31] | 1064 | 1024×436 | Small and large |
| FlyingThings3D [38] | 21818 | 960×540 | Small and large |
| Flying Chairs [12] | 22782 | 384×512 | Small and large |

Table 1: Commonly used optical flow datests.

Almost all existing datasets are limited to large rigid-body motions or quasi-rigid motions and deal with natural images. Hence, an appropriate dataset consisting of deformed speckle images had to be developed in the present study. This dataset should be made of pairs of images mimicking typical reference and deformed speckles, as well as their associated deformation fields. It has to be representative of real images and deformations so that the resulting CNN has a chance to perform well when inferring the deformation field from a new pair of images that are not in the training dataset. Speckle images have a smaller variability than natural images processed by computer vision applications. However, we shall see that we cannot use smaller datasets as the one of Table 1 because we seek tiny displacements.

*4.2. Developing a speckle dataset*

In order to estimate very small subpixel displacements (of the order of $10^{-2}$ pixel), we propose a new dataset called Speckle dataset[1]. Two versions were developed in this study. The first one, referred to as Speckle dataset 1.0 in the following, contains 36,663 pairs of frames (this number is justified below) with their corresponding subpixel displacements fields. A second version called Speckle dataset 2.0 was also developed, as explained later in this paper. Speckle dataset 1.0 was generated as follows:

1. Generating speckle reference frames.
2. Defining the displacement fields.
3. Generating the deformed frames.

These different steps are detailed in turn in the following subsections.

*Generating reference speckle frames.* In real experiments, the first step is to prepare the specimen, often by spray-painting the surface in order to deposit black droplets on a white surface. This process was mimicked here by using the speckle generator proposed in [39]. This generator consists in randomly rendering small black disks superimposed in an image, in order to get synthetic speckled patterns that are similar to actual ones. Reference frames of size $256 \times 256$ pixels were rendered with this tool. This frame size influences the quality of the results, as discussed in Section 6.1. These reference frames were obtained here by mixing the different settings or parameters which are listed in Table 2 (see [39] for more details).

|  | Speckle datasets 1.0 and 2.0 | Star images |
| --- | --- | --- |
| Probability distribution function of the radius of the disks | Uniform, Exponential and Poisson | Exponential |
| Average radius of disks | 0.45 to 0.8 pixel by step of 0.025 | 0.5 |
| Average number of disks per image | 36,000 | 556,667 |
| Contrast of the speckle | 0.5 to 1 by step of 0.05 | 0.6 |
| Size of the images | $256 \times 256$ | $501 \times 2000$ |
| Average number of disks per pixel | 0.549 | 0.556 |

Table 2: Parameters used to render the images for Speckle dataset 1.0 and 2.0 (second column), and the images deformed through the Star displacement of Section 5.3 (third column).

The quality of these reference frames was visually estimated and those featuring very large spots or poor contrast were eliminated. Only 363 frames were kept for the next step. Figure 2-a shows a typical speckle obtained with our generator and used to build Speckle dataset 1.0. It can be visually checked that the aspect is similar to a real speckle used for DIC in experimental mechanics, see typical examples in [1].

---

[1]Reference speckle frames used in speckle dataset generation and Matlab code are available at github.com/DreamIP/StrainNet

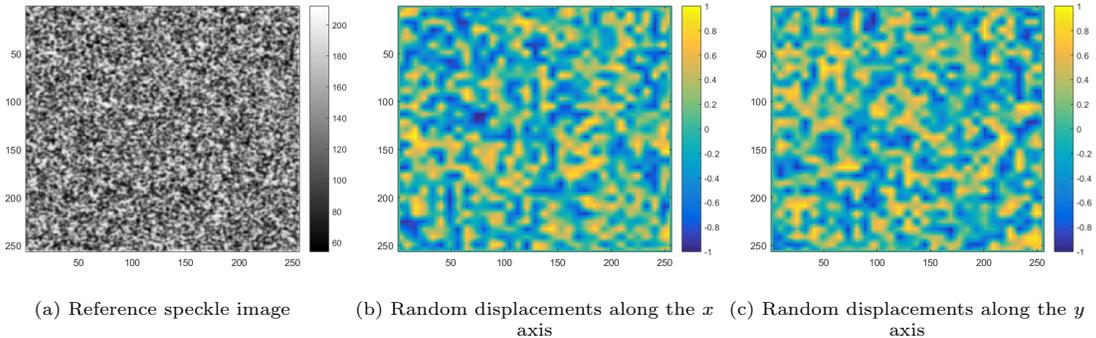

(a) Reference speckle image  (b) Random displacements along the $x$ axis  (c) Random displacements along the $y$ axis

Figure 2: Typical synthetic speckle images and random displacement (in pixels) along $x$ and $y$ used in Speckle dataset 1.0. All dimensions are in pixels.

*Defining the displacements.* The following strategy was used in order to generate a rich set of displacement fields covering the widest possible types of displacement fields. The first step consisted in splitting each reference frame into adjacent square regions of size $8 \times 8$ pixels. The pixel located at the corners of these regions were affected by a random displacement. The second step consisted in linearly interpolating the displacements between all these pixels to get the displacement at the remaining pixels of each square region.

Since we were interested here in estimating subpixel displacements, the imposed random displacements lied between -1 and +1 pixel. Furthermore, the displacements at pixels located along the boundary were set to zero to limit the error due to the missing information in these regions. Figures 2-b and -c show typical random displacements used to deform reference speckle images to obtain their deformed counterparts.

*Generating the deformed frames.* Each of the 363 reference frames was deformed 101 times, each time by generating a new random displacement field. $363 \times 100 = 36,300$ pairs of frames were used to form the dataset on which the network was trained. 363 other pairs were used for validation purposes in order to assess the ability of the network to render faithful displacement fields after training. As in other studies dealing with optical flow estimation, this ability is quantified by calculating the so-called Average Endpoint Error (AEE). AEE is the Euclidean distance between predicted flow vector and ground truth, averaged over all the pixels. Thus

$$\text{AEE} = \frac{1}{KL} \sum_{i=1}^{K} \sum_{j=1}^{L} \sqrt{(u_e(i,j) - u_g(i,j))^2 + (v_e(i,j) - v_g(i,j))^2} \quad (1)$$

where $(u_e, v_e)$ and $(u_g, v_g)$ are the estimated optical flow and the ground truth, respectively, defined at each pixel coordinate $(i, j)$. $K$ and $L$ are the dimensions of the zone over which the AEE value is calculated. This quantity is in pixel, the displacements being expressed in pixel.

The main interest of the speckle generator described in [39] is that it does not rely on interpolation when deforming the artificial speckled patterns in order to prevent additional biases caused by interpolation. However, since it relies on a Monte Carlo integration scheme, a long computing time is required. This generator is therefore only suitable for the generation of a limited number of pairs of synthetic reference and deformed images. Such an approach is not adequate here since several thousands of images had to be generated. The solution adopted here was to use bi-cubic interpolation to deform the reference images through the randomly-defined displacement fields, considering that it yielded a reasonable trade-off between computing time and quality of the results. Hence the speckle generator described in [39] was only employed to generate the reference frames mimicking real speckles used in experimental mechanics.

# 5. Fine-tuning networks of the literature

In general, CNNs are developed for specific applications and trained with datasets suited to the problem to be solved. The question is therefore to know if the studies reviewed in Section 3 above, which have good performance for estimating large displacements, can be fully or partially used to address the problem of the present paper, namely, resolving subpixel ($\sim 0.01$ pixel) displacements. Transfer learning and fine tuning are possibilities to respond to this question, as explained in Section 2.

When classification is the ultimate goal, transfer learning is carried out by replacing only the last fully-connected layer and training the model by updating the weights of the replaced layer, while maintaining the same weights for the convolutional layers. The idea behind this is that if the dataset at hand is similar to the training dataset, the output of the convolutional layers keeps on being relevant for the problem of interest. Since our dataset is completely different from the datasets used for training the CNNs described in the studies reviewed above, we proceeded to a fine tuning by updating the weights of all the layers and not only the last one. The weights found in the cases discussed above were considered as starting points for the fine-tuning performed on our own dataset.

In the following, we discuss the fine tuning of the networks considered as the most interesting ones in terms of accuracy and computational efficiency. We did not choose to fine-tune Flownet 2.0 because of its high computational cost, as discussed in Section 3. Instead, we mainly relied on the FlowNet, PWC-NET, and LiteFlownet models [12, 30, 29] because they exhibit better computational efficiency during both fine tuning and inference. In practice, we used the PyTorch implementations of these models provided in [40], [41], and [42], respectively.

## 5.1. Training strategy

The following strategy was used in order to fine-tune the networks proposed in [40, 41, 42]:

1. Keeping the network architecture, loss function, and optimization method of the original model.
2. Increasing the batch size to 16 in order to speed up the training process on our hardware configuration.
3. Initiating the learning rate (the specific machine learning vocabulary is defined in Section 2 above) with $\lambda = 10^{-4}$, and then dividing this rate by 2 every 40 epochs. These values correspond to a good trade-off between computing time and accuracy of the results.
4. Fine-tuning each network for 300 epochs because this is a value for which the value of the loss function of the considered models stabilizes.

With this strategy, fine-tuning each of the networks on an NVIDIA TESLA V100 GPU required between three to four days, depending on the model complexity.

## 5.2. Obtained results

The AEE value defined in Equation 1 was used to evaluate the fine-tuning results of the different networks. It was averaged over the test dataset made of 363 displacement fields deduced from the 363 pairs of reference and deformed speckle images considered for validation purposes. Figure 3 gives the average AEE value for each network trained on the Speckle dataset 1.0. The main conclusion is that the average AEE value is much lower with FlowNetS, which hints that more accurate estimations are expected with this network. It has also been observed that the average AEE was about the same for both the train and the test datasets for all the networks under consideration (for this reason, Figure 3 only reports the AEE of the test dataset), suggesting that they do not suffer from overfitting. Finally, the main conclusion is that the AAE value is equal at best to 0.1 pixel, which is too high since a resolution of 0.01 pixel is expected to be obtained to reach a similar performance to DIC. Since fine-tuning was not sufficient, these preexisting networks had to be improved, as explained in Section 6.

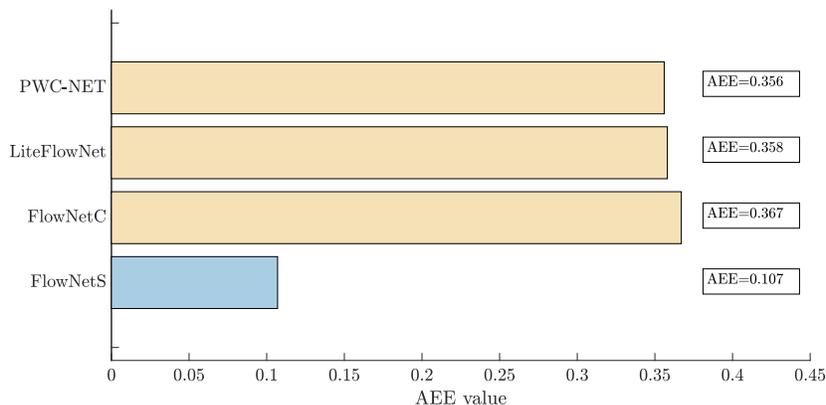

Figure 3: Average AEE value for four networks trained on Speckle dataset 1.0.

*5.3. Assessing the results with a reference Star displacement field*

Another synthetic displacement field than the interpolated random field described above has been proposed in recent papers to assess the metrological performance of various full-field measurement methods used in experimental mechanics [43, 7, 44, 10, 4]. This displacement field, named "Star displacement" in the following to be consistent with the name given in [4], is a synthetic vertical displacement modelled by a sine wave, whose period gently and linearly increases when going toward the right-hand side of the map. This vertical reference displacement field $v_g$ is shown in Figure 4b ($u_g$ is null in this case). In this figure, the green rectangle is the zone over which the results are numerically evaluated. It is not centered to emphasize the influence of the high spatial frequencies and to account for some practical constraints imposed by the use of LiteFlowNet and PWC-NET. The speckle images deformed through this displacement field are named "Star images". They were obtained by using the generator described in [39]. It means that no interpolation was performed, which was not the case for the deformed images of the dataset. The parameters used to render these speckle images are given in Table 2, third column. It can be seen that the average number of disks per pixel is nearly the same for all types of images.

The benefit of using images deformed with this displacement field is that the effect of the spatial frequency on the measured displacement field can be easily assessed by observing the attenuation of the displacement amplitude when going toward the left-hand side of the map. In other words, this reference displacement field permits the characterization of the transfer function of the estimation process. Images deformed through this displacement field were therefore considered in this study to validate the different variants of the network we developed, and more precisely to observe the attenuation of the displacement rendered for the highest spatial frequencies located on the left.

A pair of reference image and image deformed through the Star displacement field[2] corresponding to the reference displacement field was used to evaluate the results obtained with those given by the four fine-tuned CNNs selected above, namely FlowNetS, FlowNetC, LiteFlownet and PWC-NET. In the following, extension "-ft" is added to the names of the networks when they are fine-tuned. No artificial noise was added to these images. The obtained results were compared to those given by classic subset-based DIC. The latter was performed with two subset sizes, namely $2M + 1 = 11$ and 21 pixels, where $M$ is an integer which governs the size of the subset. Interpolation was performed by using spline functions. First-order subset shape functions, which are the most common ones in commercial codes [1], were considered here ( "subset shape functions" are also called "matching functions" in the literature, but we adopt here the terminology proposed in [45]). Note that results obtained with second-order subset shape functions are also presented and discussed in Section 7. As in recent papers dealing with the metrological performance of full-field measurement systems [10, 46], the shift between two consecutive subsets is equal to one pixel, which puts us in a position where DIC is at its best performance level (regardless of computing time). The results obtained in these different cases are given in Figure 5.

---

[2]Star frames used here are available at `github.com/DreamIP/StrainNet`

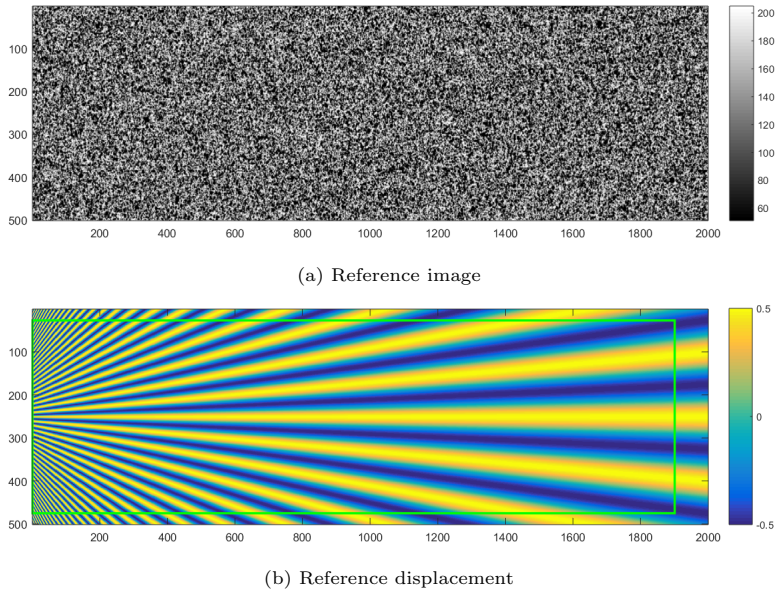

(a) Reference image

(b) Reference displacement

Figure 4: (a) Reference image corresponding to (b) the Star displacement. The green rectangle is used in this work to evaluate the results. All dimensions are in pixels.

The worst results are obtained with FlownetC. Those obtained with LiteFlownet and PWC-NET are better, but it is clear that FlownetS provides the best results, with a displacement map which is similar to that obtained with DIC and $2M + 1 = 11$ pixels. The poor accuracy obtained by the first three networks (FlownetC, LiteFlownet, and PWC-Net) may be due to the fact that these networks make use of predefined blocks such as correlation or regularization, and that they were originally developed for the determination of large displacements instead of subpixel ones.

As in previous studies dealing with quality assessment of full-field measurement techniques, such as [7, 46, 4] for instance, the displacement along the horizontal axis of symmetry, denoted here by $\Delta$, is plotted in Figure 6 in order to clearly see the attenuation of the signal when going to the highest spatial frequencies of the Star displacement (thus when going to the left of the displacement map), as well as the high-frequency fluctuations affecting the results. It is worth remembering that the reference value of the displacement along $\Delta$ is equal to 0.5 pixel. A blue horizontal line of ordinate 0.5 pixel is therefore superimposed to these curves to visualize this reference value. The closer the curves to this horizontal line, the better the results. Displacements obtained with DIC are also superimposed for comparison purposes. The mean absolute error obtained column-wise is also given.

It is worth noting that at high spatial frequencies (left-hand part of the graphs), FlowNetS provides results which are similar to those given by DIC with a subset of size $2M + 1 = 11$ pixels. At low spatial frequencies (right-hand part of the graphs), DIC provides smoother and more accurate results than FlowNetS. Calculating in each case the mean absolute error for the vertical displacement gives an estimation of the global error over the displacement map in each case. This quantity is calculated over the green rectangle plotted in Figure 4b to remove boundary effects. This quantity, denoted by MAE in the following, is defined by the mean value of the absolute error throughout the field of interest. Thus

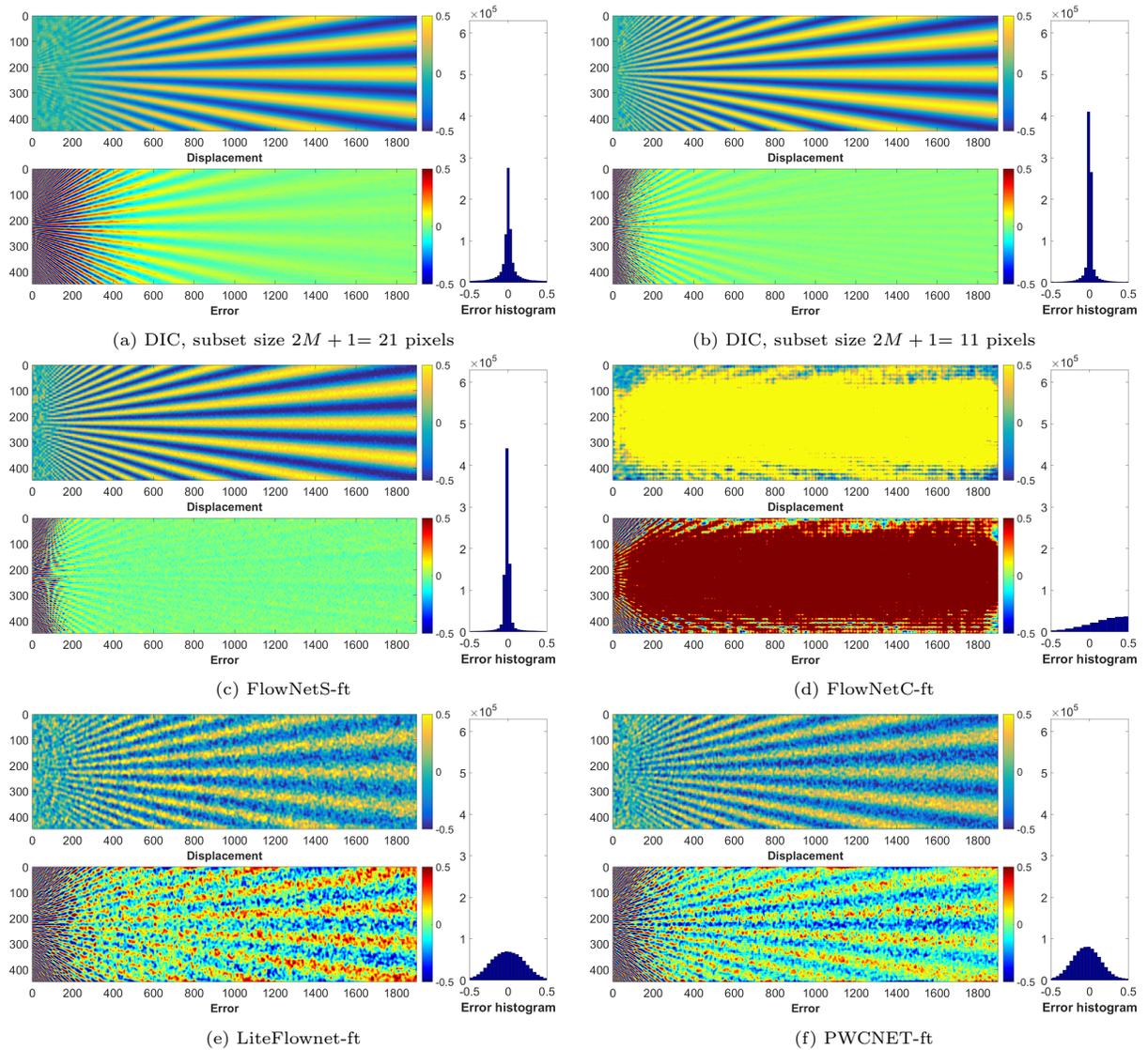

Figure 5: Star displacement obtained (a)(b) with DIC and (c)-(f) with four selected CNNs. In each case, the retrieved displacement field, the difference with the reference displacement field and the histogram of this difference are given in turn in the different sub-figures. All dimensions are in pixels.

$$\text{MAE} = \frac{1}{KL} \sum_{i=1}^{K} \sum_{j=1}^{L} |v_e(i,j) - v_g(i,j)| \qquad (2)$$

where $v_e$ and $v_g$ are the estimated displacement and ground truth, respectively. $K$ and $L$ are here the dimensions of the green rectangle over which the results are calculated. MAE is equal to AEE defined in Equation 1, in which the horizontal displacement has been nullified. It is introduced here in order to focus on the error made along the vertical direction only, the relevant information being along this direction for the displacement maps extracted from the Star images. Table 3 gives this quantity calculated for DIC and FlowNetS-ft.

It is clear that the MAE value is the lowest for DIC with $2M+1=11$ pixels. It is followed by FlowNetS-ft. These first results are promising but they come from predefined networks which are therefore not specifically developed to resolve the problem at hand. The best candidate, namely FlowNetS-ft, was therefore selected for further improvement. This is the aim of the next section.

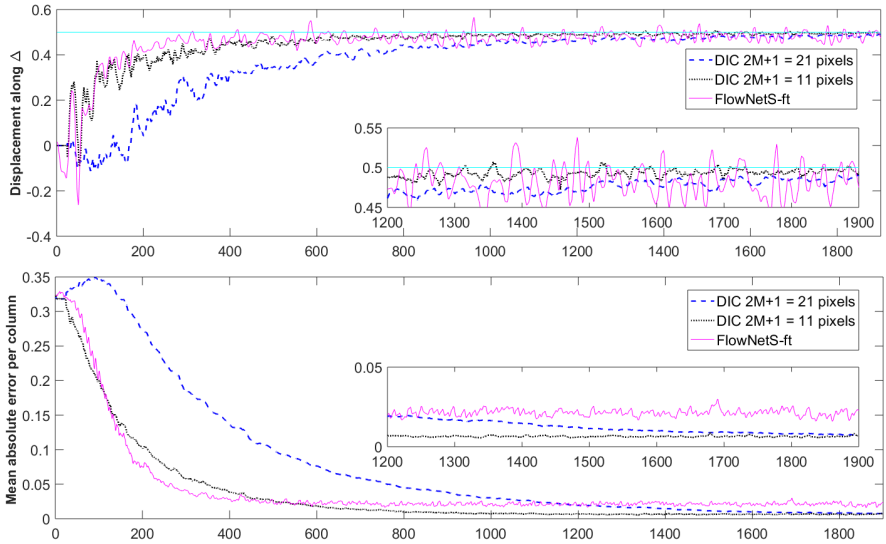

Figure 6: Comparison between FlowNetS-ft and DIC. Top: displacement along the horizontal axis of symmetry of the Star displacement field. The horizontal blue line corresponds to the reference displacement along $\Delta$. This reference displacement is equal to 0.5 pixels. Bottom: mean absolute error obtained column-wise. Close-up views of the rightmost part of the graphs are also inserted. All dimensions are in pixels.

| Technique used | DIC, $2M+1 = 21$ pixels | DIC, $2M+1 = 11$ pixels | FlowNetS-ft |
|---|---|---|---|
| MAE | 0.0828 | 0.0365 | 0.0437 |

Table 3: MAE (in pixels) for DIC (subset size $2M+1 = 11$ and 21 pixels) and FlowNetS-ft.

## 6. Tailoring FlowNetS to estimate displacement fields

Two types of modifications were proposed to enhance the accuracy of the results. The first one concerns the network, the second the dataset.

### 6.1. Improvement of the network

First, let us examine in more details the architecture of FlowNetS [12]. This network can be split into two different parts, as illustrated in Figure 1. The first one extracts the feature maps using 10 successive convolutional layers. The first layer uses $7 \times 7$ filters, the second and third layers use $5 \times 5$ filters, and the remaining layers use $3 \times 3$ filters. The second part predicts the optical flow at different levels and relies, in the case of FlowNetS, on 5 convolutional layers with $3 \times 3$ filters and 8 transposed convolutional layers. The transposed convolutional layers are used for the up-sampling process (see the right-hand side of the schematic view depicted in Figure 1). We refer the interested reader to [12] for more details on the architecture.

As mentioned in Section 3, the FlowNetS architecture provides an optical flow of 1/4 of the original image resolution. The full-resolution optical flow is obtained by applying an up-scaling step with a factor of 4 using a bi-linear interpolation. This not only reduces the computational complexity but also the quality of the results. In order to enhance the prediction of the optical flow, it was therefore proposed to improve the network to directly output a higher resolution optical flow, without interpolation.

In general, the deeper the network, the better the results. The dataset shall however be enlarged accordingly to avoid overfitting (see definition in Section 3 above)). Two approaches were examined in this study in order to enhance the accuracy of the predicted optical flow. The first one consists in adding some levels to the network (thus increasing the number of layers), the second in changing the architecture.

*6.1.1. First approach: adding one or two levels*

The first approach consists in adding levels to the network. A first option consists in adding one level to the FlowNetS network while keeping the same architecture. Consequently, the new output optical flow has a resolution divided by 2 instead of 4. This method still requires interpolation in order to reach the same resolution as the input images. A second option has also been examined. It consists in adding one more level so that the full-resolution is directly obtained without any interpolation.

The loss function defined in Section 2 is equal to the following quantity

$$Loss = \sum_{i=1}^{n} \lambda_i e_i \quad (3)$$

where $n$ is the number of levels forming the network, $\lambda_i$ is a weighting coefficient corresponding to the $i$-th level, and $e_i$ is the AEE between of the output of the $i$-th level and the ground-truth displacement sampled at the same resolution. This loss function was adapted to the proposed networks by keeping the same $\lambda_i$ coefficients as in [12] for the levels corresponding to the FlowNetS levels, and affecting a coefficient of 0.003 to each new level. Compared to the strategy defined in Section 5.1, only the loss function is modified. Two training scenarios were used here. In the first one, the new networks were fine-tuned by updating all the weights of the networks. In the second scenario, only the weights of the new levels were updated. The remaining weights were the same as those obtained after applying the fine-tuning process described in the previous section.

By applying the first scenario for training the new network with one additional level only, the average AEE value defined in Section 5.2 increases from 0.107 to 0.141, which means that the results are worse. The same trend is observed with the MAE value deduced from the displacement map obtained with the Star images (MAE = 0.3334). The displacement map obtained in this case is shown in Figure 7. It can be seen that the error made is significant. Adding one more level and retraining all the network weights does not improve the quality of the results, which means that the first scenario is not a good option for training the networks.

On the contrary, the second scenario really improves the accuracy of the results when considering the random displacement fields used to generate the 363 deformed speckle images considered as test dataset, with a reduction of the average AEE of more than 50% compared to FowNetS-ft, as reported in Table 4. The average AEE concerns the 363 displacement fields retrieved from the 363 pairs of images of the test dataset. When considering now the Star displacement, we can see that the MAE reported in Table 4 for each of these two new networks is nearly the same as the MAE obtained with FlowNetS-ft. This is confirmed by visually comparing the Star displacement reported in Figure 8 which are obtained with these two new networks, and the reference Star displacement depicted in Figure 4b. Besides, it can be seen in Figure 8 that almost the same accuracy as FlowNetS-ft is obtained when considering the Star displacement field of Figure 4b.

The displacements along the horizontal axis of symmetry $\Delta$ of the Star displacements which are obtained with these two networks, FlowNetS-ft, and DIC ($2M + 1 = 11$ and $2M + 1 = 21$ pixels) are depicted in Figure 9 to more easily compare the different techniques. No real improvement is clearly observed with these curves. The mean absolute error estimated column-wise is lower for the proposed networks than for DIC ($2M + 1 = 11$ pixels) for medium-range spatial frequencies, between $x = 200$ and $x = 400$, but this error becomes higher for $x > 600$. The obtained results show that the networks

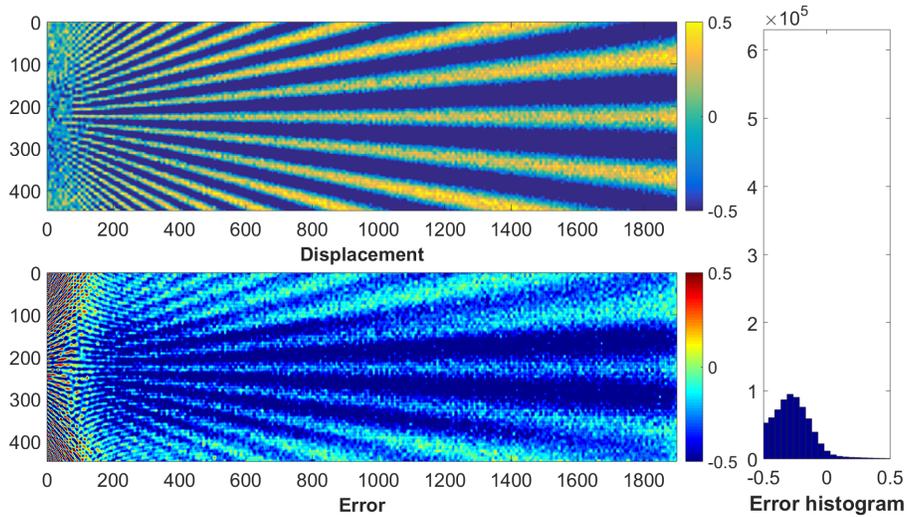

Figure 7: Star displacement obtained by adding one level only to FlowNetS, performing interpolation to reach full resolution and updating all the weights, (MAE = 0.3334). All dimensions are in pixels.

| Metric | FlowNetS-ft | DIC, $2M+1 = 11$ pixels | Network with one additional level | Network with two additional levels |
|---|---|---|---|---|
| Average AEE | 0.1070 | - | 0.0560 | 0.0450 |
| MAE | 0.0437 | 0.0365 | 0.0445 | 0.0471 |

Table 4: Comparison between *i*- FlowNetS-ft, *ii*- DIC with $2M+1 = 11$ pixels and *iii*- FlowNetS after adding one or two levels and updating only the weights of the new levels. Average AEE calculated over the whole images of the test dataset and MAE calculated with the Star displacement.

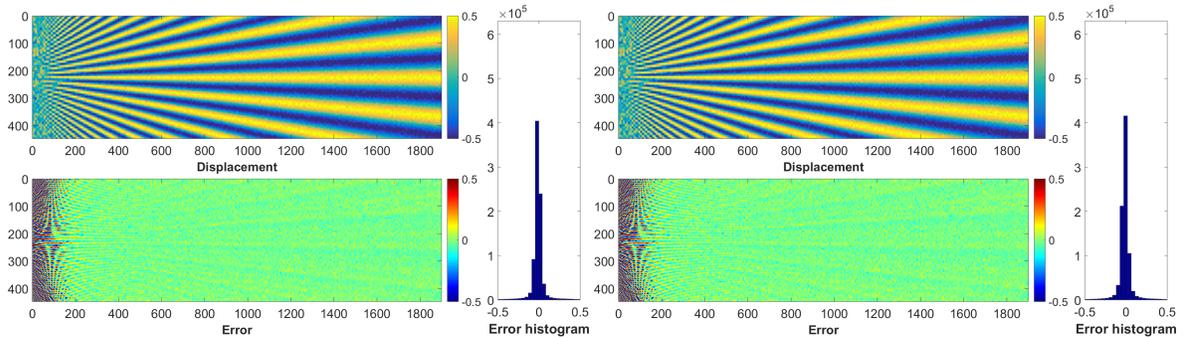

(a) Option 1: Adding one level to FlowNetS, MAE = 0.0445.

(b) Option 2: Adding two levels to FlowNetS, MAE = 0.0471.

Figure 8: Star displacement obtained by adding one or two levels to FlowNetS and updating only the weights of the new levels. All dimensions are in pixels.

proposed here enhance the learning accuracy on Speckle dataset 1.0 compared to FlowNetS-ft but no improvement is observed with the Star displacement.

*6.1.2. Second approach: simplifying the architecture of FlowNetS*

In FlowNetS, the feature maps are down-sampled 6 times (6 strides of 2 pixels), which means that the resolution of the last feature maps is $2^6$ smaller than the resolution of the images which are processed. Then, optical flow estimation and up-sampling processing are performed only 4 times, which leads to output an optical flow with a resolution divided by 4. We also propose two alternative networks with this second approach. The first one, named StrainNet-f, is a full-resolution network obtained by applying 4 down-samplings followed by 4 up-samplings. The second network, named StrainNet-h, is a half-resolution network obtained by applying 5 down-samplings followed by 4 up-samplings. The same loss function as in FlowNetS is used in both cases. These two networks are trained by using the weights of the corresponding levels of FlowNetS-ft as starting values, and then by fine-tuning all the network weights. The same training strategy as that described in Section 5.1 was adopted. The average AEE and MAE values reported in Table 5 clearly show that these two proposed networks outperform the previous ones, even though no real difference is visible to the naked eye between the Star displacements reported in Figure 10 and those reported in Figure 8. This is clearer when observing the displacements along $\Delta$ reported in Figure 11 and comparing them with their counterparts in Figure 9. Indeed the sharp fluctuations visible in the closeup view embodied in Figure 11 are much smaller and smoother than those shown in Figure 9. In addition, the MAE per column is lower in the latter than in the former at high frequencies (for about $200 < x < 400$). Finally, the main conclusion of the different metrics given in Tables 4 and maps or curves showed in Figure 10 and 11 is that the last two networks perform better than DIC ($2M + 1 = 11$ and 21 pixels, 1st order) at high spatial frequencies, and that they provide comparable results at low frequencies. Let us now examine how to improve further the results by changing the dataset used to train the networks.

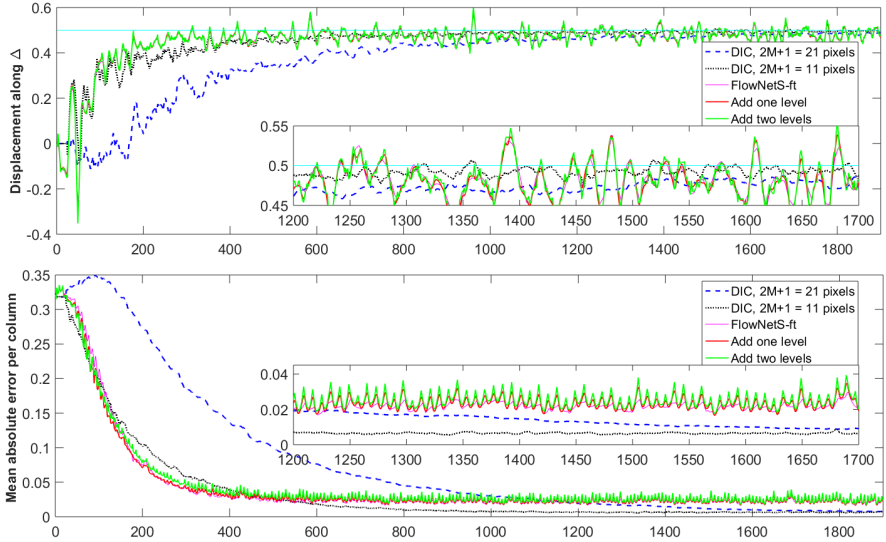

Figure 9: Comparison between *i*- FlowNetS-ft, *ii*- FlowNetS after adding one or two levels and updating only the weights of the new levels, and *iii*- DIC with $2M + 1 = 11$ and $2M + 1 = 21$ pixels. Displacements along $\Delta$ and mean absolute error per column.

| Metric | FlowNetS-ft | DIC, $2M + 1 = 11$ pixels | Half-resolution network | Full-resolution network |
|---|---|---|---|---|
| Average AEE | 0.1070 | - | 0.0352 | 0.0211 |
| MAE | 0.0437 | 0.0365 | 0.0350 | 0.0361 |

Table 5: Comparison between *i*- FlowNetS-ft, *ii*- DIC with $2M + 1 = 11$ pixels, and *iii*- FlowNetS after changing the architecture with two options: half-resolution (StrainNet-h) and full resolution (StrainNet-f), and updating all the weights. Average AEE and MAE (in pixels), calculated over the whole images of the test dataset.

Different methods were investigated in this section in order to improve the network. Table 6 gathers all these methods and the corresponding results in order to help the reader compare them and easily find the results obtained in each case. The results can also be improved by changing the dataset. This is the aim of the next section.

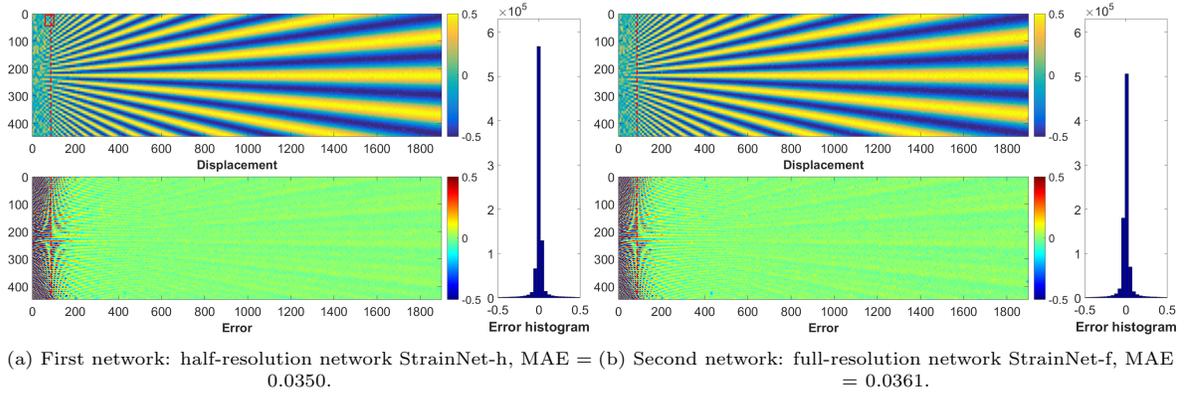

(a) First network: half-resolution network StrainNet-h, MAE = 0.0350.

(b) Second network: full-resolution network StrainNet-f, MAE = 0.0361.

Figure 10: Star displacement obtained after changing the architecture of FlowNetS with two options: (a) half-resolution and (b) full resolution, and updating all the weights. All dimensions are in pixels. The abscissa of the vertical red line is such that the period of the sine wave is equal to 16 pixels, thus twice the size of the regions used to define the displacement maps gathered in Speckle dataset 1.0. The red square at the top left is the zone considered for plotting the closeup view in Figure 12.

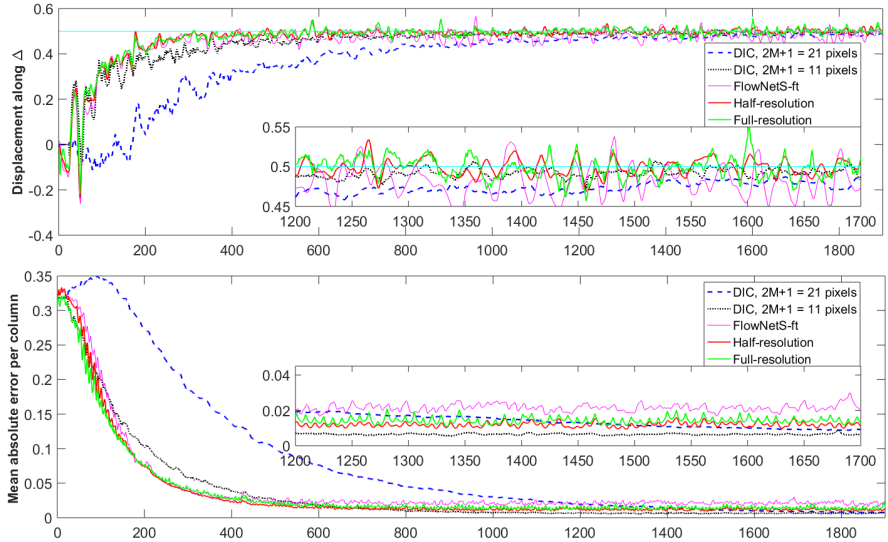

Figure 11: Comparison between *i*- FlowNetS-ft, *ii*- DIC with $2M + 1 = 11$ and 21 pixels, and *iii*- FlowNetS after changing the architecture with two options: half-resolution and full resolution, and updating all the weights. Displacements along $\Delta$ and mean absolute error per column. All dimensions are in pixels.

### 6.2. Improvement of the training dataset

#### 6.2.1. Procedure

An interesting conclusion of the preceding simulations is that, regardless of the influence of sensor noise propagation to final displacement maps, the networks proposed in the first approach (see Section 6.1.1) enhance the average accuracy over the test dataset. Nevertheless, the accuracy of the Star displacement field returned by these networks is not improved, especially when high-frequency displacements are concerned. This suggests limited generalization capability of the proposed network, probably caused by the nature of the training dataset.

We now discuss how to change the speckle dataset so that a lower noise and a lower bias can be obtained with both StrainNet-f and -h for the highest spatial frequencies of the Star displacement.

| Approach | Option | Training scenario | Results |
|---|---|---|---|
| 1- Adding Levels | one level | Updating all the weights | Fig. 7 |
| 1- Adding Levels | one level | Updating new weights only | Table 4, Fig. 8a |
| 1- Adding Levels | two levels | Updating new weights only | Table 4, Fig. 8b |
| 2-Changing architecture | Half-resolution (StrainNet-h) | Updating all the weights | Table 5, Fig. 10a, Fig. 11 |
| 2- Changing architecture | Full-resolution (StrainNet-f) | Updating all the weights | Table 5, Fig. 10b, Fig. 11 |

Table 6: Methods and options investigated to improve the network in this study.

*Observing the bias at high frequencies in the Star displacement map.* Before improving the dataset, let us examine in detail the results obtained in the preceding section. For instance, the error maps depicted in Figure 10 show that the error suddenly increases for the highest frequencies (those on the left). Interestingly, the location of this sudden change in response of the network substantially corresponds to the zone of the displacement map for which the spatial period of the vertical sine wave is equal to twice the size of the region used in Speckle dataset 1.0, namely $8 \times 2 = 16$ pixels. This is confirmed by superimposing to these maps a vertical red line at an abscissa which corresponds to this period of 16 pixels. The explanation is that the displacement field considered in Speckle dataset 1.0 are linear interpolation of random values 8 pixel apart. They are therefore increasing or decreasing over an 8 pixel interval, and cannot correctly represent sine waves of period lower than 16 pixels.

In the same way, let us now enlarge the displacement field obtained with StrainNet-h trained on Speckle dataset 1.0. The zone under consideration is a small square portion of the displacement field (see precise location with the red square in Figures 10a and 13a). The result is given in Figure 12a. It can be observed that the network is impacted by the fact that the dataset used for training purposes is generated from $8 \times 8$ independent regions. Indeed, the network predicts the displacement at one point per region and then interpolates the results to obtain the full optical flow. This phenomenon is confirmed by down-sampling a predicted displacement with a factor of 8 and then up-sampling the result with the same factor. The resulting displacement is practically the same as the one given by StrainNet-h trained with Speckle dataset 1.0. The main conclusion is that the network cannot correctly predict the displacements on the left of the Star displacement because they occur at higher spatial frequencies than those used in Speckle dataset 1.0.

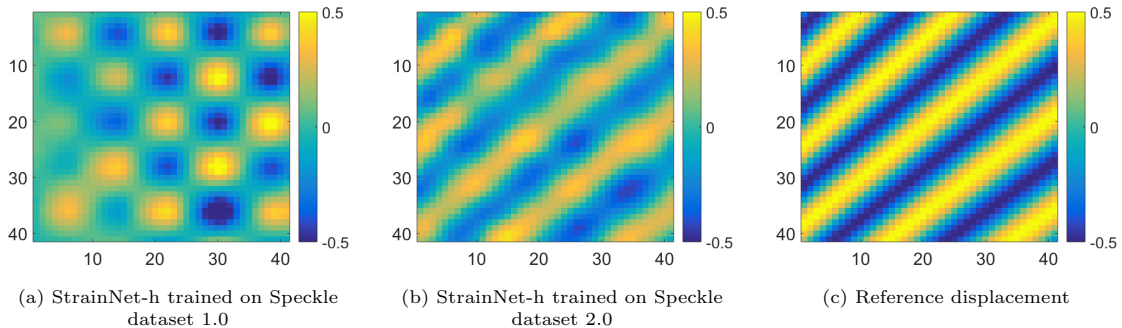

Figure 12: Results obtained with StrainNet-h trained on Speckle dataset 1.0 and 2.0: closeup view at high spatial frequencies area of the Star displacement (see precise location in Figures 10 and 13). (a) StrainNet-h trained on Speckle dataset 1.0. (b) StrainNet-h trained on Speckle dataset 2.0. (c) Reference displacement. All dimensions are in pixels.

A consequence of the remarks above is that square regions smaller in size than 8 pixels should also be included in the speckle dataset to be able to retrieve displacement fields involving spatial frequencies higher than 1/8 pixel$^{-1}$. A second and more suitable dataset called Speckle dataset 2.0 was therefore generated, as explained below.

*Generating Speckle dataset 2.0.* Speckle dataset 2.0 was generated with the same principle as Speckle dataset 1.0, but by changing three design rules. First, regions of various sizes instead of uniform size of $8 \times 8$ pixels were considered to define the random ground truth displacement fields. On the one hand, the preceding remark motivates the use of smaller regions. On the other hand, less accurate estimation than with DIC for low-frequency displacements (see for instance Figure 11) motivates the use of larger regions. We therefore considered regions of size equal to $4 \times 4$, $8 \times 8$, $16 \times 16$, $32 \times 32$, $64 \times 64$ and $128 \times 128$ pixels.

Second, bilinear interpolation used in Speckle dataset 1.0 was replaced by a bicubic one to limit potential interpolation bias. Third, a noise was added to all the images of the dataset in order to simulate the effect of sensor noise which always corrupts digital images, while only noiseless images were considered in Speckle dataset 1.0. This noise was heteroscedastic to mimic typical sensor noise of actual linear cameras [47, 43]. With this model, the variance $v$ of the camera sensor noise is an affine function of the brightness $s$, so $v = a \times s + b$. We chose here a=0.0342 and b=0.2679 to be consistent with the values used to generate the noisy images used in the DIC Challenge 2.0 [4].

The number of frames was the same for each region size. It was equal to $363 \times 10 = 3,630$ (363 reference images deformed 10 times). Since six different region sizes were considered, Speckle dataset 2.0 contains $6 \times 3630 = 21780$ different pairs of images.

*6.2.2. Results obtained with the Star images*
*Results obtained with noiseless images.* StrainNet-h and StrainNet-f were trained again, but this time on Speckle dataset 2.0 instead of Speckle dataset 1.0. Processing then the noiseless Star images with these two networks gives the results illustrated in Figure 13-a and -b, respectively. It is worth remembering that StrainNet-f anf -h are used here to determine a displacement field from noiseless frames after being trained on noisy frames.

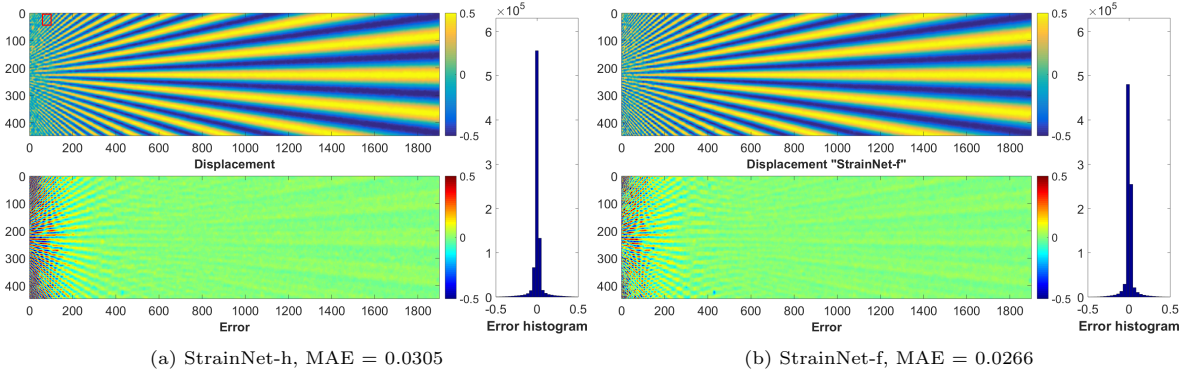

(a) StrainNet-h, MAE = 0.0305     (b) StrainNet-f, MAE = 0.0266

Figure 13: Star displacement obtained with StrainNet-f and StrainNet-h trained on speckle dataset 2.0. The red square at the top left is the zone considered for plotting the closeup view in Figure 12. All dimensions are in pixels.

Results obtained at the right-hand side of the displacement map of Figure 13 are rather smooth. In addition, bearing in mind that the colorbars used in Figures 10 and 13 are the same, it can be seen by comparing the error maps that the high spatial frequencies are rendered in a more accurate way with the networks trained on Speckle dataset 2.0., in particular the high-frequency components in the left-hand side of the displacement maps.

Figure 14 compares the results obtained by StrainNet-h and StrainNet-f trained in turn on Speckle datasets 1.0 and 2.0. The results obtained by these networks trained on Speckle dataset 2.0 are also compared in Figure 15 with DIC (subset size $2M + 1 = 11$ and 21 pixels). It is clear that the results obtained after training on Speckle dataset 2.0 are better and have less fluctuations of the error. Furthermore, the results shown in Figure 15 show that StrainNet-h and StrainNet-f have a similar accuracy to DIC at low frequencies and a better one at high frequencies.

Finally, it can be concluded that training the network with Speckle dataset 2.0 instead of Speckle dataset 1.0 leads to better results with the Star images. Let us now examine the influence of image noise on the results.

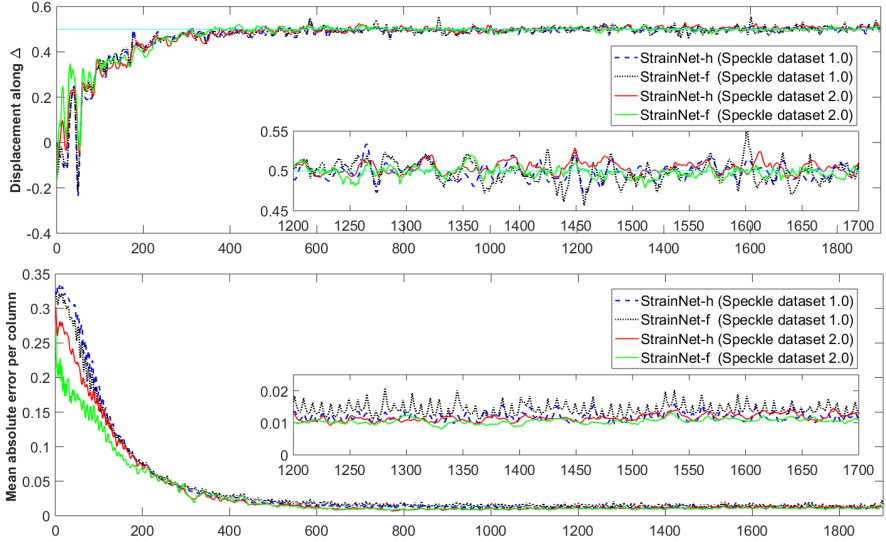

Figure 14: Comparison between the networks trained on Speckle dataset 1.0 and Speckle dataset 2.0.

*Results obtained with noisy images.* The previous results were obtained with noiseless images. We now consider noisy images to evaluate the dependency of the maps provided by StrainNet-f and -h to image noise. The Star images used in this case were obtained by adding a heteroscedastic noise similar to the one discussed above to the noiseless Star images.

Figure 16 shows the maps obtained with StrainNet-h and StrainNet-f with the Star images, as well as those obtained with DIC for comparison purposes ($2M + 1 = 11$ and 21 pixels). It is worth noting that StrainNet-h and StrainNet-f outperform DIC $2M + 1 = 11$ pixels at both high and low spatial frequencies. Indeed, the bias is lower for the high spatial frequencies and a smoother displacement distribution (thus a lower noise) is obtained.

## 7. Spatial resolution and metrological performance indicator

### 7.1. Spatial resolution

As in other studies dealing with the metrological performance of full-field measurement techniques [10, 46], the spatial resolution of each algorithm is estimated here by considering the displacements along the horizontal axis of symmetry $\Delta$. The reference value is 0.5 pixels all along this line, but the measured one becomes lower when going toward the high frequencies of the star displacement, thus toward the left of the map. The attenuation of the signal, which directly reflects the bias, depends on the spatial frequency. The spatial resolution, denoted here by $d$, is generally defined as the inverse of the spatial frequency obtained for a given attenuation of the amplitude of the vertical sine displacement. This attenuation is generally equal to 10%=0.1. In the present case, it means that the spatial resolution is the period of the vertical sine displacement for which the amplitude is equal to $0.5 - 0.5 \times 0.1 = 0.45$ pixel. The value of $d$ must be as small as possible to reflect a small spatial resolution, thus the ability of the measuring technique of interest to distinguish close features in displacement and strain maps, and return a value of the displacements and strains in these regions with a small bias. In certain cases, the displacement resolution can be predicted theoretically from the transfer function of the filter associated to the technique (Savitzky-Golay filter for DIC [5, 6], Gaussian filter for the Localized Spectrum Analysis [48]).

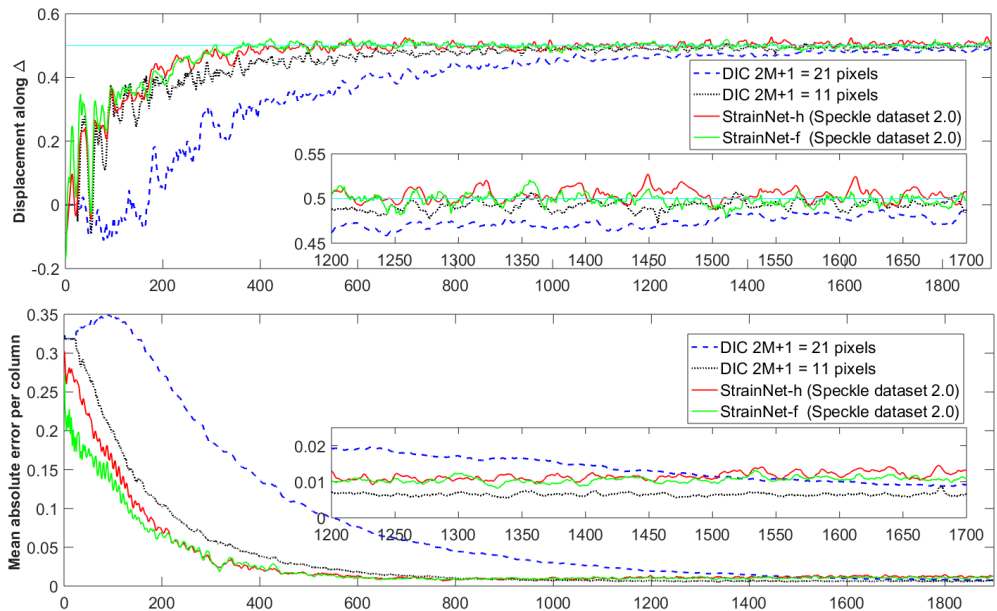

Figure 15: Comparison between the networks trained on Speckle dataset 2.0 and DICs ($2M+1 = 11$ and 21 pixels). Top: displacements along $\Delta$. Bottom: mean absolute error per column. All dimensions are in pixels.

In the present case of CNN however, no transfer function has been identified so far, so $d$ can only be determined numerically or graphically, by seeking the intersection between the curve representing the error obtained along $\Delta$ with noiseless images on the one hand, and a horizontal line of equation $y = 10\%$ on the other hand, see Figure 17. Note that the curves were smoothed with a rectangular filter to remove the local fluctuation of the bias that could potentially disturb a proper estimation of this quantity. The spatial resolution found in each case is directly reported in each subfigure. We also considered here second-order subset shape functions, this case being more favorable for DIC [49]. Only the case $2M+1 = 21$ pixels is reported here, DIC diverging at some points with $2M+1 = 11$ pixels.

The value of $d$ is smaller with both StrainNet-f and -h than with DIC used with first-order subset shape functions, even for $2M+1 = 11$ pixels, while $d$ is nearly the same with DIC used with second-order subset shape functions. Indeed, Figure 18 (top) shows that the displacement along $\Delta$ is similar between StrainNet-f or -h on the one hand, and DIC with second-order subset shape functions on the other hand. Figure 19, where closeup views of the error map for the highest spatial frequencies are depicted, shows however that the way $d$ is estimated is more favorable for DIC with second-order subset shape functions than for StrainNet-f and -h. Indeed, an additional bias occurs with DIC when going vertically away from the axis of symmetry $\Delta$, as clearly illustrated in Figure 19a. Figures 19b-c show that it is not the case for StrainNet-f and -h. This phenomenon is not taken into account when estimating $d$ since only the loss of amplitude along $\Delta$ is considered. Consequently, when considering the mean absolute error per column as in Figure 18 (bottom), it can be observed that this error is

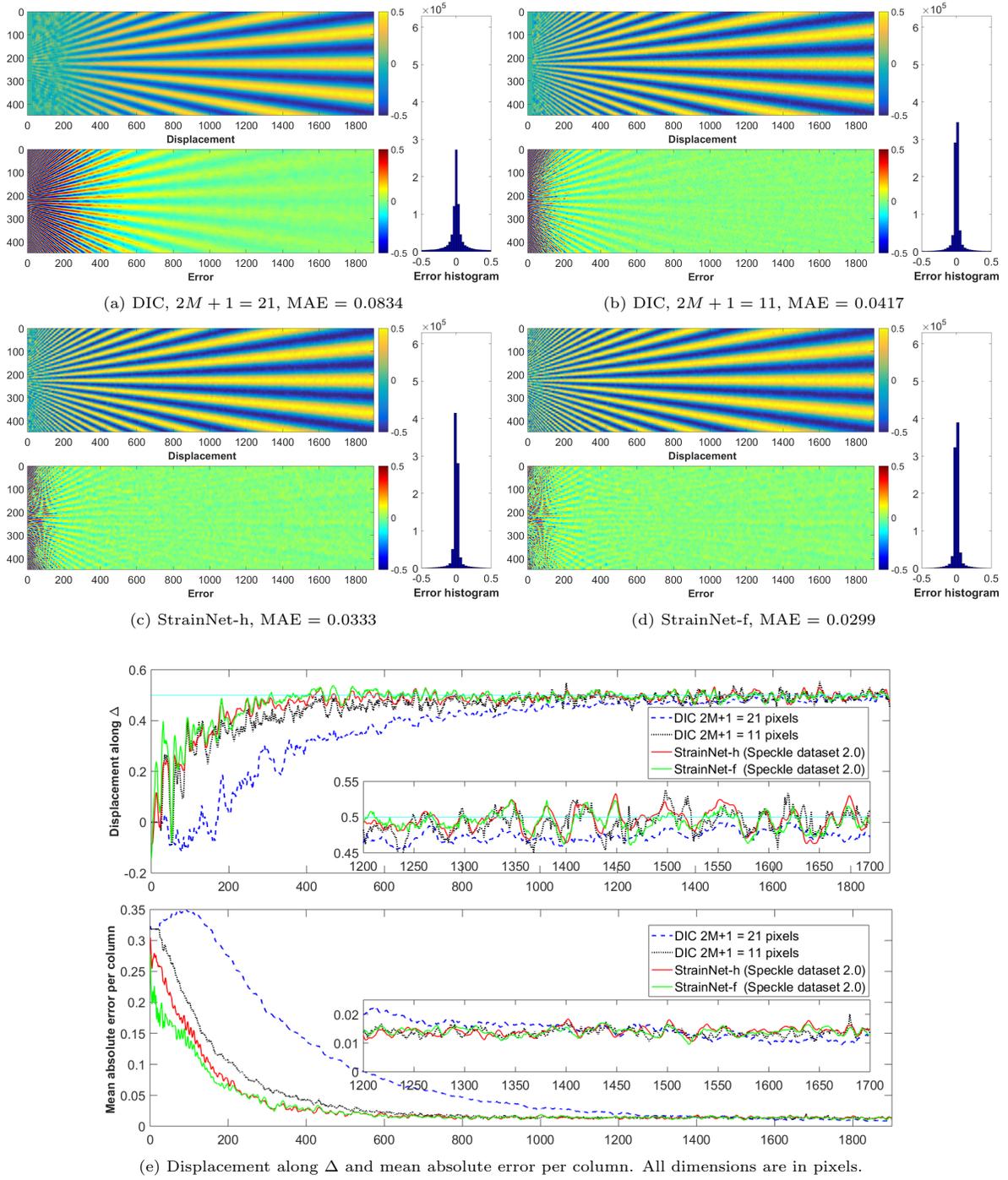

(a) DIC, $2M+1 = 21$, MAE = 0.0834

(b) DIC, $2M+1 = 11$, MAE = 0.0417

(c) StrainNet-h, MAE = 0.0333

(d) StrainNet-f, MAE = 0.0299

(e) Displacement along $\Delta$ and mean absolute error per column. All dimensions are in pixels.

Figure 16: Results given by DIC and StrainNet with noisy images.

lower for StrainNet-f and -h than for DIC with second-order subset shape functions.

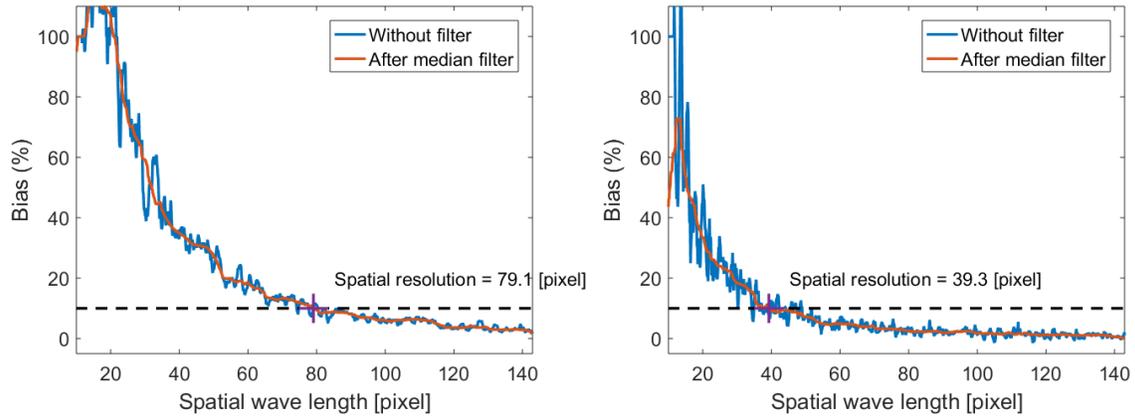

(a) DIC, $2M + 1 = 21$ pixels, first-order subset shape functions (b) DIC, $2M + 1 = 11$ pixels, first-order subset shape functions

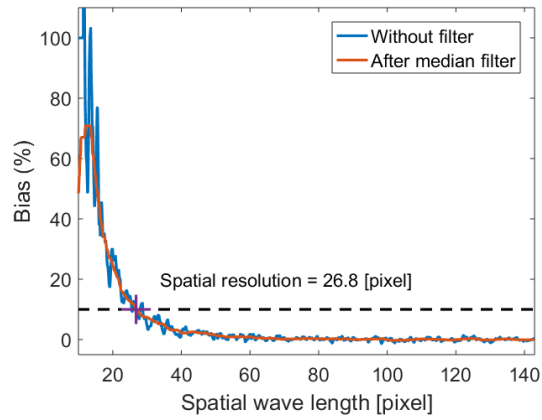

(c) DIC, $2M + 1 = 21$ pixels, second-order subset shape functions

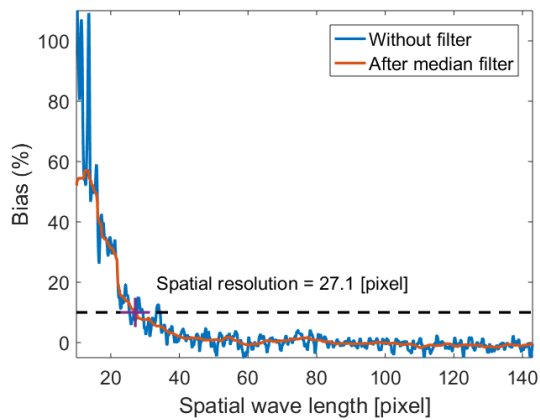
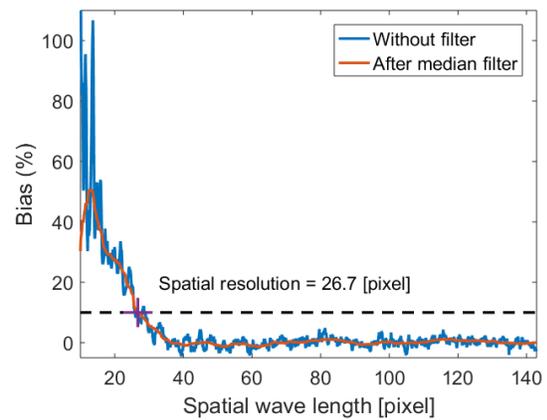

(d) StrainNet-h  (e) StrainNet-f

Figure 17: Seeking the spatial resolution of each technique. The bias given here is a percentage of the displacement amplitude, which is equal to 0.5 pixel
.

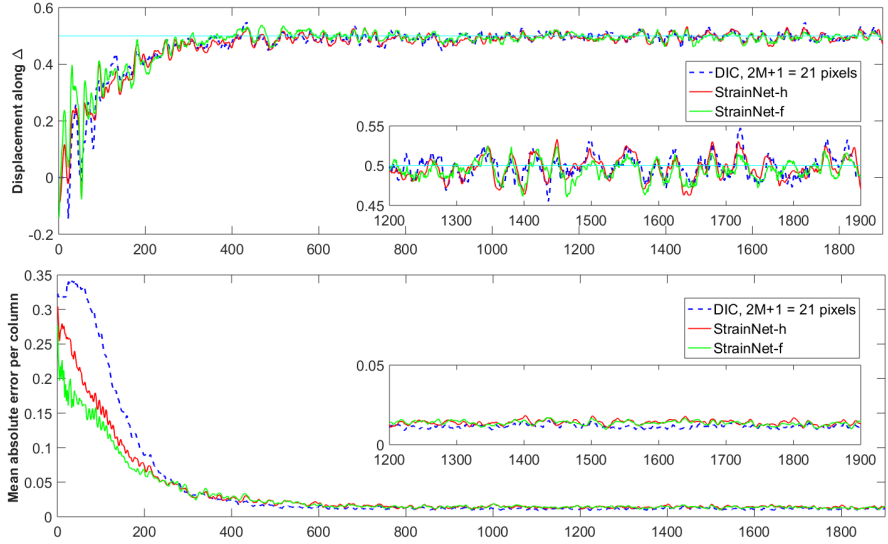

Figure 18: Comparison between StrainNet and DIC ($2M + 1 = 21$ pixels) with second-order subset shape function (noisy images). Top: displacements along $\Delta$. Bottom: mean absolute error per column. All dimensions are in pixels.

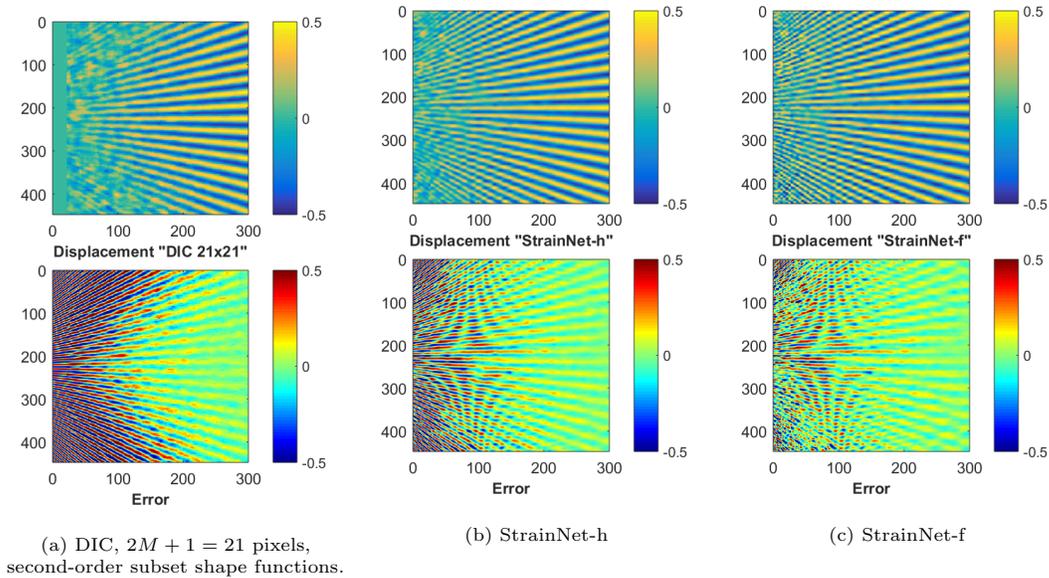

(a) DIC, $2M + 1 = 21$ pixels, second-order subset shape functions.

(b) StrainNet-h

(c) StrainNet-f

Figure 19: Closeup view of the error map in pixels (for the high spatial frequencies) obtained with StrainNet and DIC with second-order subset shape functions.

### 7.2. Metrological performance indicator

A general remark holds for full-field measurement systems: the lower the value of $d$, the higher the noise in the displacement map. The noise level is quantified by the standard deviation of the noise. This quantity, denoted by $\sigma_u$, reflects the displacement resolution of the technique. Thus there is a link between $d$ and $\sigma_u$. It has been rigorously demonstrated in [48, 43] that this product is constant for the Localized Spectral Analysis, a spectral technique which minimizes the optical residual in the Fourier domain. It has also been observed in subsequent studies that it was also the case for DIC [7, 46], which minimizes the same residual, but in the spatial domain. Hence estimating the product between $d$ and $\sigma_u$ is a handy way to compare measurement systems for a given bias, but independently from any other parameter chosen by the user such as the subset size for DIC. This product, denoted by $\alpha$ and named "metrological performance indicator" in [7, 46], has been calculated here for the different algorithms. The value of $\sigma_u$ is merely estimated by calculating the standard deviation of the difference between the displacement fields found by processing noisy and noiseless Star images. The values of $\alpha$ found for each algorithm is reported in Figure 20.

This indicator is nearly the same for DIC used with $2M + 1 = 11$ pixels and 21 pixels (1st order), which is consistent with the conclusions of [7]. It is also almost identical for StrainNet-f and -h. Both lie between DIC used with first- and second-order subset shape functions. Since the spatial resolution estimated with $d$ is nearly the same with DIC used with second-order subset shape functions and StrainNet, it means that the noise level is higher in the latter case. This can be observed in the noise maps depicted in Figure 21. In particular, the shape of the wave can be guessed in Figure 21b-c. A higher difference can also be observed on the left, so for the highest spatial frequencies. It means that a slight bias is induced by noise in these two cases. Further investigations should be undertaken to see how to eliminate this phenomenon, by changing the dataset and/or the layers of the network itself.

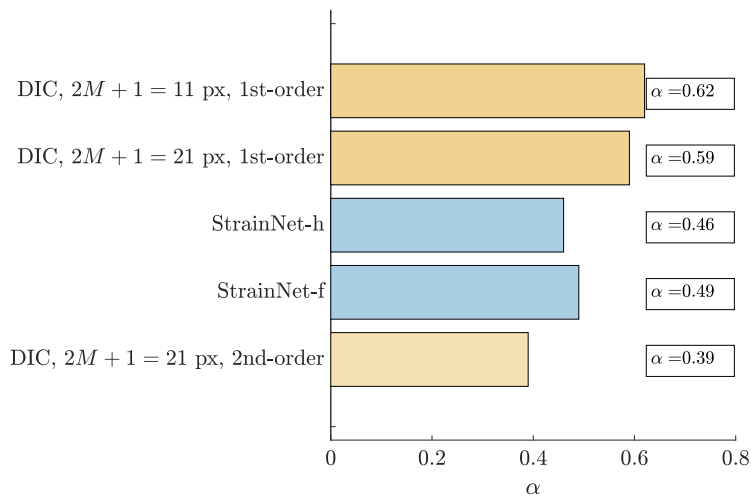

Figure 20: Metrological efficiency indicator $\alpha$ for DIC (1st and 2nd subset shape functions), StrainNet-h and StrainNet-f.

7.3. Pattern-induced bias

Pattern-induced bias (PIB) is a phenomenon, which has only recently been observed and described in the DIC literature [50, 51]. This bias is induced by the pattern texture itself. It manifests itself by the presence of random fluctuations in the displacement and strain maps. These random fluctuations shall not be confused with sensor noise propagation since it is due to different causes, mainly the image gradient distribution and the difference between the true displacement field and its local approximation by subset shape functions, see [6] where a model for this phenomenon is proposed. These spatial fluctuations are randomly distributed because speckle patterns are randomly distributed. The aim here is to briefly examine whether displacement fields retrieved by StrainNet are also prone to this phenomenon. We performed for this the same two experiments as in [51]. These experiments also rely on synthetic images deformed through the Star displacement. The first one consists in considering a unique pair of reference/deformed speckle images, adding 50 times a different copy of heteroscedastic noise, retrieving the 50 corresponding displacement fields, and plotting the mean distribution along the horizontal axis of symmetry $\Delta$ of the displacement map. The second one consists in considering 50 different pairs of reference/deformed speckles images, adding heteroscedastic noise to each image,

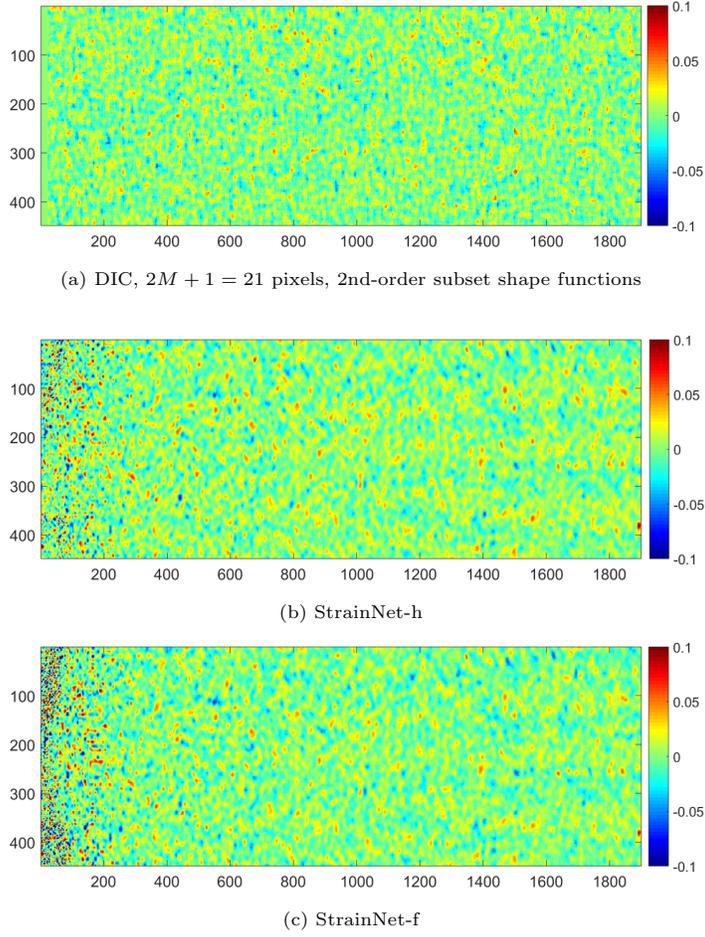

(a) DIC, $2M + 1 = 21$ pixels, 2nd-order subset shape functions

(b) StrainNet-h

(c) StrainNet-f

Figure 21: Difference between displacement fields obtained with noisy and noiseless speckle images (in pixels). All dimensions are in pixels.

retrieving the 50 corresponding displacement fields, and plotting again the mean distribution along $\Delta$. With the first experiment, the random fluctuations due to sensor noise are averaged out (or at least decrease in amplitude by averaging effect). However, PIB is constant over the dataset since the displacement is the same and the speckles are similar from one image to another, the only difference between them being due to noise. In the second experiment, all the speckles are different and they are noisy, so both the random fluctuations due to sensor noise on the one hand, and due to the random fluctuations cause by PIB on the other hand, are averaged out. Comparing the curves obtained in each of these two cases enables us to numerically illustrate the effect of PIB on the displacement field, and to study its properties.

Figure 22 shows on the left and on the right the curves obtained on average in the first and second cases, respectively. They are plotted in red. The curves obtained for the 50 different pairs of images are superimposed in each case to give an idea of the fluctuations which are observed with each pair of images. The results obtained with DIC ($2M + 1 = 11$ pixels, 1st order and $2M + 1 = 21$ pixels, 2nd order) are also shown for comparison purposes. It is worth remembering that exactly the same sets of images are processed here by DIC and StrainNet. The main remark is that PIB also affects the results given by StrainNet, but averaging the results provided by 50 different patterns less improves the results than for DIC. The effect of PIB seems therefore to be less pronounced for StrainNet than for DIC, other phenomena causing these fluctuations. It is worth remembering that StrainNet-f and -h were trained on Speckle dataset 2.0, and that the deformed images contained in this dataset were obtained by interpolation. It would therefore be interesting to train StrainNet-f and -h with images obtained with the Speckle generator described in [39]. Indeed this latter generator does not perform any interpolation, so we could see if the errors observed in Figure 22-f are due to the pattern alone or to both the pattern and the bias due to the interpolation used when generating the deformed images. A larger dataset with another random displacement generation scheme as in Speckle dataset 2.0 could also help smoothing out this bias.

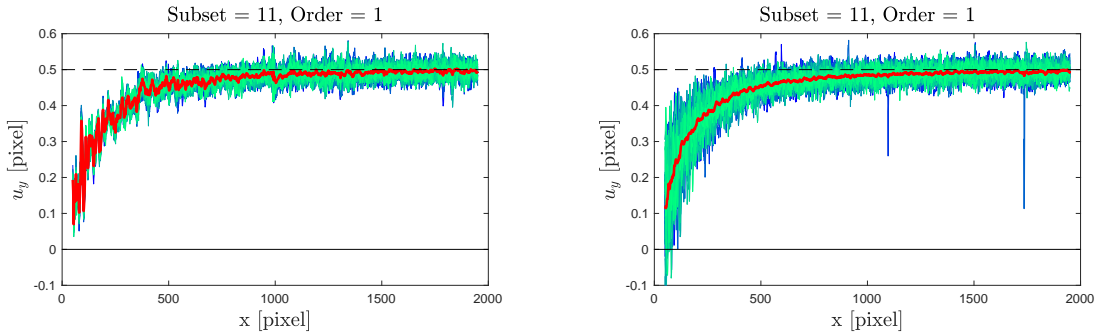

(a) DIC, $2M + 1 = 11$ pixels, 1st-order subset shape functions, 1 pattern with 50 noises

(b) DIC, $2M + 1 = 11$ pixels, 1st-order subset shape functions, 50 patterns with a different noise

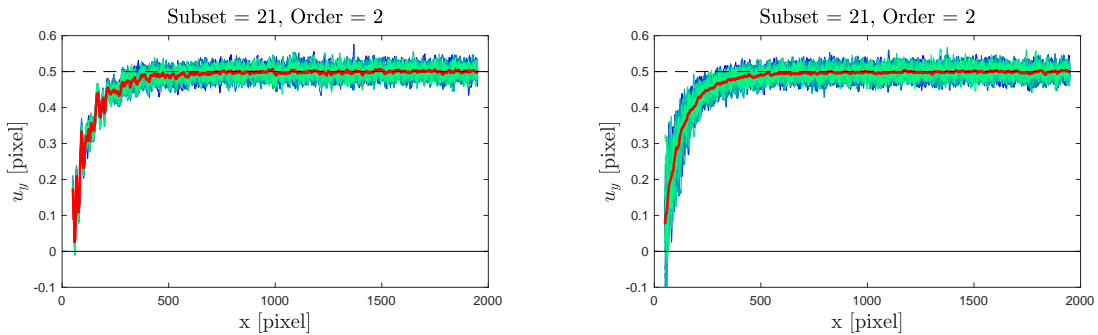

(c) DIC, $2M + 1 = 21$ pixels, 2nd-order subset shape functions, 1 pattern with 50 noises

(d) DIC, $2M + 1 = 21$ pixels, 2nd-order subset shape functions, 50 patterns with a different noise

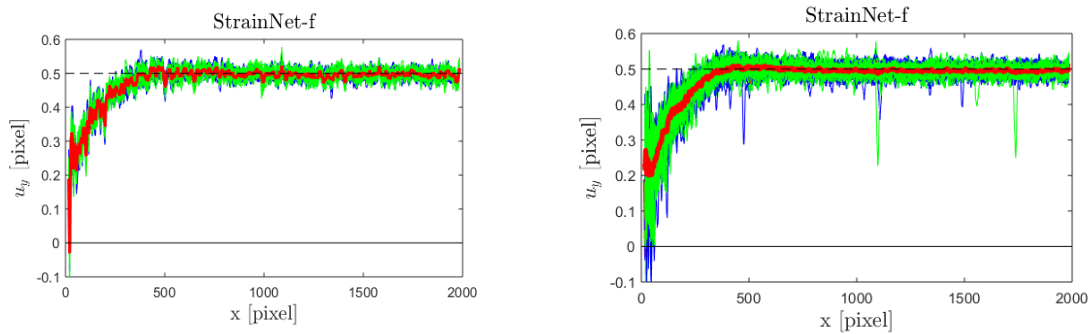

(e) StrainNet-f, 1 pattern with 50 noises

(f) StrainNet-f, 50 patterns with a different noise

Figure 22: Pattern-induced bias. Comparison between results obtained with DIC ($2M + 1 = 11$ pixels, 1st order and $2M + 1 = 21$ pixels, 2nd order) and StrainNet-f.

## 8. Assessing the generalization capability

In deep learning, a key point is to validate any CNN with images different from those used to train the network in order to ensure good generalization capability. In the preceding sections, we took care to use StrainNet on speckle images deformed with the Star displacement while this network was trained with Speckle dataset 2.0 which does not contain any image deformed in a similar way. This is, however, not sufficient because both the reference images in Speckle dataset 2.0 and the reference Star image were generated with the same speckle generator [39]. Two other examples were therefore considered in this study. Both involve speckles, which are different from those obtained with the speckle generator [39] which generates the reference frames in the speckle datasets, and both the reference and deformed Star images. The first example concerns images of synthetic speckles from Sample 14 of the DIC Challenge 1.0 [4], the second real speckle images taken during a compression test performed on a wood specimen, as described in [44].

### 8.1. Example 1: Sample 14 of the DIC Challenge 1.0

In this section, we consider a pair of images from Sample 14 of the DIC Challenge 1.0 [4, 3]. The speckle pattern is obtained with TexGen, a speckle generator described in [52]. This pattern is deformed by using a standard FFT expansion. It can be checked that the visual aspect is different from that of the images used so far in this paper. In particular, large dots over which the image gradient is nearly null can be observed, see Figure 23a where a typical subset size of $2M + 1 = 21$ pixels is superimposed. For comparison purposes, Figure 23b represents a closeup view of one the speckle images of Speckle dataset 2.0. The displacement used to artificially deform the images is a sine wave along the horizontal axis, but with a frequency which increases when going to the right, the amplitude being constant and equal to 0.1 pixel. This example is discussed in [3], so the same colorbar as in this reference is used here to make comparison easier. Figures 23c-j show the displacement fields $u$ along the horizontal direction obtained with the images from the "Sample 14 L5 Amp0.1" file available in [4]. DIC ($2M + 1 = 11$ and 21 pixels, first-order subset shape functions), StrainNet-h, and -f are used here. The mean value along the vertical direction of the horizontal displacement, denoted here by $\overline{u}_x$, is also represented.

It can be seen that results obtained with DIC and $2M + 1 = 21$ pixels is the less affected by noise. Noise increases with $2M + 1 = 11$ pixels, which is logical since the subset size is smaller. The displacement field obtained with StrainNet-h is similar to the one obtained with DIC and $2M + 1 = 11$ pixels. Figure 24 shows the strain fields $\varepsilon_{xx}$ corresponding to the displacement fields shown above. They were obtained by convolving the corresponding displacement fields with a derivative kernel, which enables us to perform at the same time both a smoothing of the noisy data and differentiation. This kernel is the $x$-derivative of a Gaussian window of standard deviation equal to 6 pixels. Again, similar maps are obtained with StrainNet-f and -h trained on Speckle dataset 2.0 on the one hand, and DIC run with $2M + 1 = 11$ pixels on the other hand. However, the main conclusion here is not really that StrainNet-f and -h give displacement and strain maps similar to DIC, but that these networks trained with speckles obtained with the speckle generator described in [39] are able to determine the displacement field used to deform speckle images obtained with another generator.

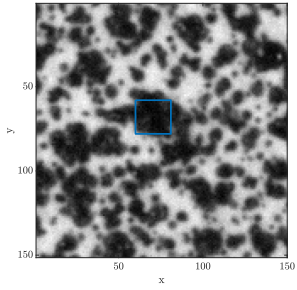

(a) Close-up view of the images of Sample 14 of the DIC challenge 1.0. A subset used in DIC is superimposed (size: $2M+1 = 21$ pixels).

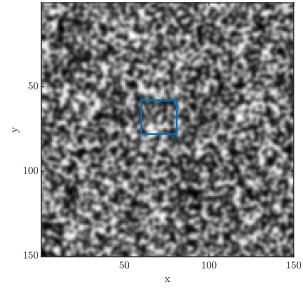

(b) Close-up view of one of the images of Speckle dataset 2.0 used to train the network.

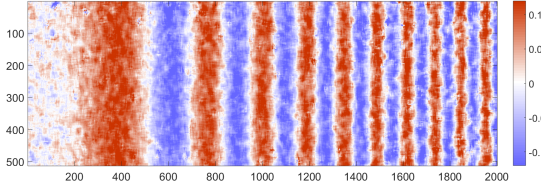

(c) $u_x$, DIC, $2M+1 = 21$ pixels

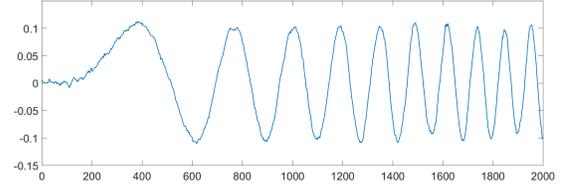

(d) $\overline{u}_x$, DIC, $2M+1 = 21$ pixels

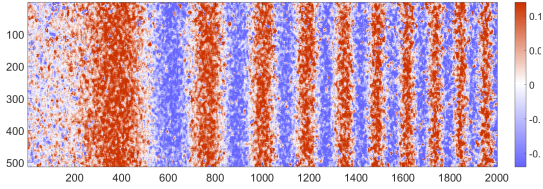

(e) $u_x$, DIC, $2M+1 = 11$ pixels

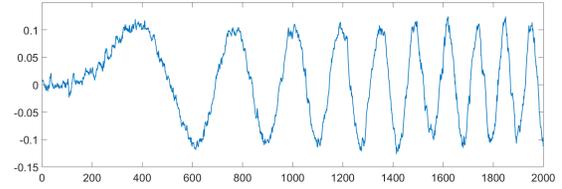

(f) $\overline{u}_x$, DIC, 11, $2M+1 = 11$ pixels

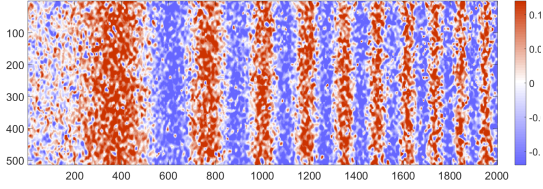

(g) $u_x$, StrainNet-h

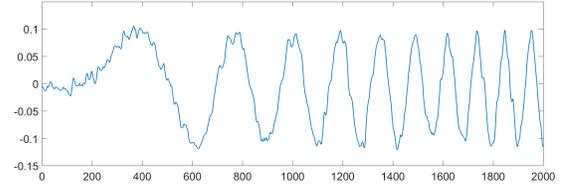

(h) $\overline{u}_x$, StrainNet-h

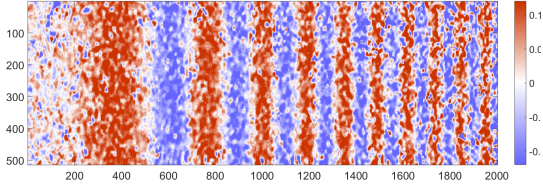

(i) $u_x$, StrainNet-f

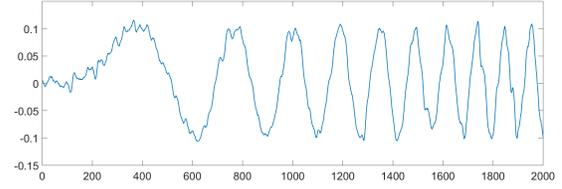

(j) $\overline{u}_x$, StrainNet-f

Figure 23: Results obtained by processing images from Sample 14 of the DIC Challenge [4]. Top: closeup view of the speckle. Left: displacement fields. Right: profile of the average displacement over all the columns of the maps. See [3] to compare these maps to those obtained with other DIC packages. All dimensions are in pixels.

## 8.2. Example 2: Compression test on a wood specimen

In this second example, we consider a real test performed on a wood specimen shown in Figure 25a, see [44] for more details about this specimen and the testing conditions. The interest here is twofold. First real speckles instead of artificial ones are considered. Second, the constitutive material of the specimen is heterogeneous since this is a stack of early and late wood. A consequence is that the stiffness spatially changes, and so does the strain distribution if the rings are perpendicular to the loading force, which is the case here. We consider a typical pair of frames and applied StrainNet-f to determine the displacement field. Results obtained with DIC with $2M + 1 = 21$ pixels (1st-order subset shape functions) are shown for comparison purposes. A convolution with a Gaussian derivative filter is then applied in all cases to deduce the vertical strain map $\varepsilon_{yy}$.

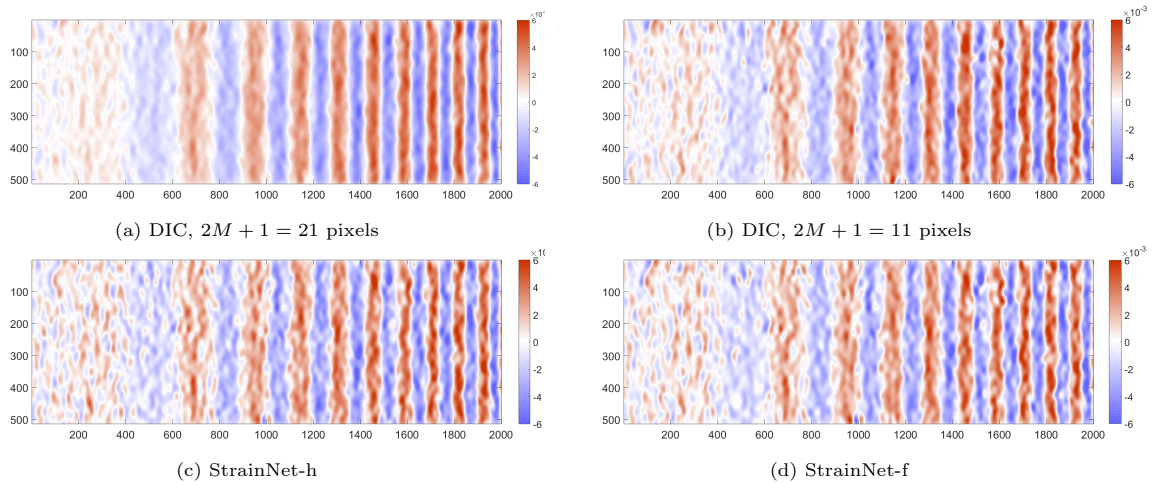

Figure 24: Strain map $\varepsilon_{xx}$ deduced from the displacement fields depicted in Figure 23. All dimensions are in pixels.

It can be seen that similar maps are obtained but again, a more refined analysis should be performed to discuss the possible damping of the actual details in the strain map, as in [44] but this is out of the scope of the present paper. The main point here is that StrainNet is able to extract displacement and strain fields featuring rather high spatial frequencies from images different from those obtained with Speckle dataset 2.0 since this is here a real speckle pattern. It must be noted that the displacement is greater than one pixel over most of the front face of the specimen. This displacement was therefore split into two quantities. The first one is the displacement with a quite rough pixel resolution. The second one is the subpixel displacement. The images are therefore processed piecewise, in such a way that the round value of the displacement is the same throughout each of the elements of the mesh. A mesh of 11 horizontal bands is considered here. This round value for the displacement can easily be found by cross-correlation for instance. This point is however not really challenging and we applied here a rough DIC to get this integer value for the sake of simplicity, only the subpixel determination of the displacement being of interest in the present study. A consequence is the fact that on close inspection, slight straight lines can be seen in the strain maps, along the border between some of the elements. Nothing can be visually detected at the same place in the displacement maps.

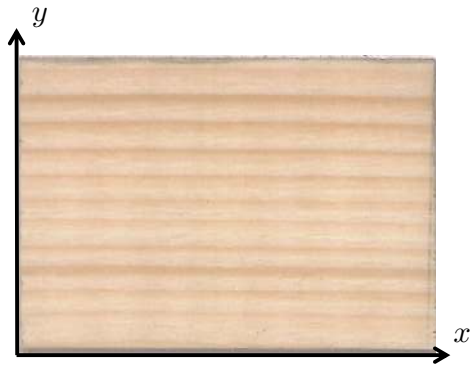
(a) Specimen before spray-painting, after [44]

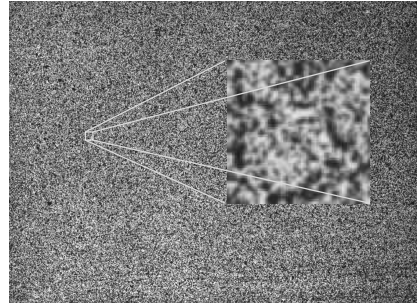
(b) Speckled surface of the specimen after spray painting, after [44]

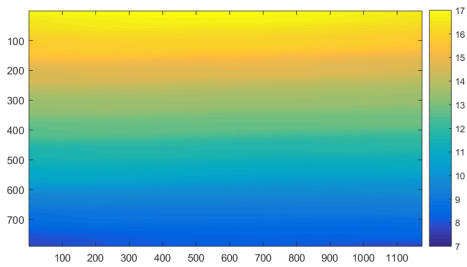
(c) DIC, $2M + 1 = 21$ pixels

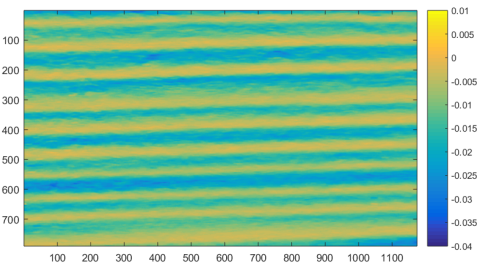
(d) DIC, $2M + 1 = 21$ pixels

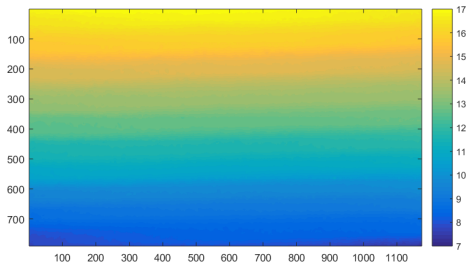
(e) StrainNet-h

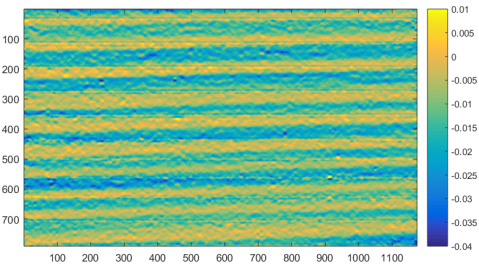
(f) StrainNet-h

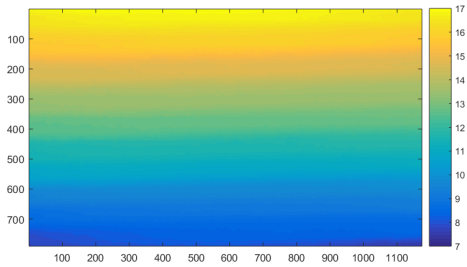
(g) StrainNet-f

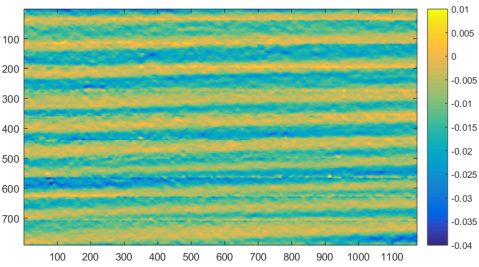
(h) StrainNet-f

Figure 25: Results obtained by processing real images. Left: $v$ displacement field in pixels. Right: $\varepsilon_{yy}$ strain map. All dimensions are in pixels.

## 9. Computing time

Finally, we give some information on the computing time needed to perform the calculations. CNNs are well suited to be run on Graphical Processing Units (GPUs), which was the case here. This is even necessary to achieve training in a reasonable amount of time, as mentioned in Section 5.1. Once StrainNet was trained, we used it on the same GPU. The computing time needed to estimate the subpixel displacement field is reported in Table 7. Two typical examples are considered here. The first one corresponds to one pair of Star images, the second to Sample 14 of the DIC Challenge 2.0. The number of pixels of the frames and the computing time are also reported in this table. The number of Points Of Interest per second (POI/s) is also given in each case. This quantity represents the number of points per second at which a measurement is provided by the measuring system. It is used in [53] in order to measure the calculating speed of GPU-based DIC. In our case, using this quantity is a handy way to normalize the results obtained with different techniques and different frame sizes, and to fairly compare them.

| Case | Frame size | StrainNet-h | | StrainNet-f | |
|---|---|---|---|---|---|
| | | Computing time (s) | POI/s | Calculating time (s) | POI/s |
| Example 1 | $2000 \times 501$ | 0.0081 | 1.24E+08 | 0.077 | 1.30E+07 |
| Example 2 | $2048 \times 589$ | 0.0088 | 1.37E+08 | 0.092 | 1.31E+07 |

Table 7: Computing time and Points Of Interest per second (POI/s) for Examples 1 and 2.

For StrainNet-f and -h, the value of POI/s is about ten times lower for the "h" version than for the "f" one while the resolution before final interpolation to reach full resolution is 4 times lower. Interestingly, Ref [53] reports a POI/s equal to $1.66 \cdot 10^5$ and $1.13 \cdot 10^5$ for a parallel DIC code implemented on a GPU, which is nearly two orders of magnitude below. These values are given for information only: the GPU used in [53] (NVIDIA GTX 760, 2.3 TFLOPs) is indeed less powerful than the GPU used in the present study (NVIDIA TESLA V100, 114 TFLOPs). In addition, the reader should bear in mind that CNNs must be trained with a suitable dataset, which generally represents heavy calculations. Further work should therefore be undertaken to fairly compare StrainNet and a GPU-based DIC in terms of computing time. The conclusion is, however, that StrainNet provides pixel-wise displacement maps (and thus strain maps by convolution with suitable derivative filters) at a rate compatible with real-time measurements.

## 10. Conclusion

This paper presented a CNN dedicated to the measurement of displacement and strain fields. As for DIC, the surface to be investigated was marked with a speckle pattern. Various strategies deduced from the similar problem of optical flow determination were presented, and the best one has been adapted to give a network named StrainNet. This network was trained with two versions of a specific speckle dataset. The main result was to demonstrate, through some relevant examples, the feasibility of this type of approach for accurate pixelwise subpixel measurement over full displacement fields. As for other problems tackled with CNNs in engineering, the main benefit here is the very short computing time to obtain the sought quantities, here the displacement fields. Further studies remain necessary to investigate various problems, which are still open after this preliminary work. For instance, the dataset used in order to train the network directly influences the quality of the final results. The dataset developed here led to valuable results for speckles different than those of the images forming the dataset, in particular experimental ones. This observation should however be consolidated by considering a wider panel of speckles and thus by trying to reduce noise and bias in the final displacement maps. The generator free from any interpolation, which was used here only for generating the deformed Star images for a matter of time, could also be employed for the images of the dataset despite the computing cost. The networks discussed here were obtained by enhancing FlowNetS. A complete redesign should also be undertaken, in particular in order to simplify the network. This would certainly reduce both the training and the processing times. Finally, a model able to deal with displacements larger than one pixel while still giving accurate subpixel estimation should also be investigated further, for instance by training the network on a dataset containing deformed images involving displacements greater than one pixel.

## Acknowledgments

This work has been sponsored by the French government research program "Investissements d'Avenir" through the IDEX-ISITE initiative 16-IDEX-0001 (CAP 20-25) and the IMobS3 Laboratory of Excellence (ANR-10-LABX-16-01).